\documentclass[runningheads]{svmult}

\usepackage{makeidx}   % allows index generation
\usepackage{graphicx}  % standard LaTeX graphics tool
\usepackage{subeqnar}  % subnumbers individual equations
\usepackage{multicol}  % used for the two-column index

\usepackage{physprbb}  % modified textarea for proceedings,
 \usepackage{amssymb}

\makeindex             % used for the subject index
  \def\bld#1{{\bf #1 }}
 \def\xb{\bar\xi(a,x)}
 \def\gaprox{\mbox{$\,$ 
\raisebox{0.5ex}{$<$}\hspace{-1.7ex}{\raisebox{-0.5ex}{$\sim$ }}$\,$} }
\def\part#1#2{\frac{\partial #1}{\partial #2}}
\def\pder#1#2{{\partial #1/\partial #2}}

\def\rb{\right)}
\def\lb{\left(}

\def\fra#1#2{{#1\over#2}}

\def\la{\mathrel{\mathchoice {\vcenter{\offinterlineskip\halign{\hfil
$\displaystyle##$\hfil\cr<\cr\sim\cr}}}
{\vcenter{\offinterlineskip\halign{\hfil$\textstyle##$\hfil\cr<\cr\sim\cr}}}
{\vcenter{\offinterlineskip\halign{\hfil$\scriptstyle##$\hfil\cr<\cr\sim\cr}}}
{\vcenter{\offinterlineskip\halign{\hfil$\scriptscriptstyle##$\hfil\cr<\cr\sim\cr}}}}}

\def\ga{\mathrel{\mathchoice {\vcenter{\offinterlineskip\halign{\hfil
$\displaystyle##$\hfil\cr>\cr\sim\cr}}}
{\vcenter{\offinterlineskip\halign{\hfil$\textstyle##$\hfil\cr>\cr\sim\cr}}}
{\vcenter{\offinterlineskip\halign{\hfil$\scriptstyle##$\hfil\cr>\cr\sim\cr}}}
{\vcenter{\offinterlineskip\halign{\hfil$\scriptscriptstyle##$\hfil\cr>\cr\sim\cr}}}}}

\begin{document}
\title*{Statistical Mechanics of gravitating systems \protect\newline in static and  cosmological backgrounds}
\toctitle{Statistical Mechanics of gravitating systems \protect\newline in static and  
cosmological backgrounds}
% allows explicit linebreak for the table of content
%
%
\titlerunning{Statistical mechanics of gravitating systems}
\author{T. Padmanbhan}
\authorrunning{T. Padmanabhan}
\institute{IUCAA, Pune University Campus, Ganeshkhind, Pune - 7, India.\\
email: nabhan@iucaa.ernet.in
}

\maketitle              

\begin{abstract}
This pedagogical review addresses several issues related to statistical description of
gravitating systems  in both static and expanding backgrounds, focusing on the latter.
 After briefly reviewing
the results for the static background, I describe the key issues and recent progress in the
context of gravitational clustering\index{gravitational clustering}  of collision-less particles in an expanding universe\index{expanding universe}.
The questions addressed include: 
(a) How does the power injected into the system at a given 
wave number  spread to smaller and larger scales?
(b) How does the power spectrum of density fluctuations\index{density fluctuations} behave asymptotically
 at late times?
(c) What are the universal characteristics  of gravitational clustering
that are independent of the initial conditions and manifest at the late time
evolution of the system?
The review is intended for non cosmologists and will be of interest to people 
working in fluid mechanics, non linear dynamics and condensed matter physics.
\index{abstract} 
\end{abstract}

  \section{\label{intro}Introduction }
  
  The statistical mechanics of systems dominated by gravity is of interest both
  from the theoretical and ``practical''  perspectives. Theoretically this field has close connections
  with areas of condensed matter physics, fluid mechanics, re-normalization group etc. and
  poses an incredible challenge as regards the basic formulation. From the practical point of
  view, the ideas find application in many different areas of astrophysics and cosmology,
  especially in the study of globular clusters, galaxies and gravitational clustering in the
  expanding universe. (For a general review of statistical mechanics of gravitating systems,
  see \cite{tppr};  textbook description of the subject is available in \cite{textone}, \cite{texttwo}. 
  Review of gravitational clustering in expanding background is available in \cite{tpiran} and
  in several textbooks in cosmology \cite{cosmotext}. There have been many
  attempts to understand these phenomena by different groups; see \cite{chavanis}, \cite{sanchez},
  \cite{vala}, \cite{fola}, \cite{botta}, \cite{roman}  and the references cited therein.)
  Given the diversity of the subject, it will be useful to begin with a broad overview
  and a  description of the issues which will be addressed in this review. 
 
  I will  concentrate mostly on the problem of gravitational clustering in the expanding universe
  which is one of the most active research areas in cosmology. However, to place this
  problem in context, it is necessary to discuss statistical mechanics of isolated gravitating
  systems (without any cosmological expansion) in some detail. In  Part I of this review,
  I cover this aspect highlighting the important features but referring the reader to existing previous
  literature for details. Part II presents a  more detailed description of gravitational
  clustering in the context of cosmology. 
  The rest of the introduction will be devoted to a  summary of different issues which will be
  expanded upon in the later sections. 
  
  Let me begin with the issues which arise in the study of isolated gravitating systems (like,
  say, a cluster of stars) treated as a collection of  structure-less point particles.
In Newtonian theory, the gravitational force can be described
as a gradient of a scalar potential and the evolution of a set of particles under the action of gravitational
 forces can be described the equations
\begin{equation}
 \ddot{\bf x}_i = - \nabla \phi ({\bf x}_i, t); \quad
\nabla^2 \phi = 4 \pi G \sum_i m_i \delta_D ({\bf x} -{\bf x}_i) 
\end{equation}
where  ${\bf x}_i$ is the position of the $i-$th particle, $m_i$ is its
mass.  For 
sufficiently large number of particles, it is useful to investigate whether 
some kind of statistical description  of such a system is possible. 
Such a description, however,  is complicated by  the  
 long range, \index{long range} unscreened, nature of 
 gravitational force.  The force acting on any given particle receives contribution even
from particles which are  far away.
 If a 
self gravitating system is divided into two parts, the total energy 
of the system cannot be expressed as the sum of the gravitational
energy of the components. Many of the 
conventional results in statistical physics are
  based on the extensivity of the 
energy \index{extensivity of the 
energy} which is clearly not valid for gravitating systems. 
 To make progress, we have to use different techniques which are appropriate for each situation. 
  
  To construct the statistical description of such a system, one should begin
  with the construction of the micro-canonical ensemble describing such a system.
  If the Hamiltonian of the system is $H(p_i, q_i)$ then the volume $g(E)$ of the constant
  energy surface  $H(p_i, q_i) =E$ will be of primary importance in the  micro-canonical ensemble\index{micro-canonical ensemble}.
  The logarithm of this function will give the entropy $S(E) = \ln g(E) $ and the temperature of the
  system will be $T(E)\equiv \beta(E)^{-1} = (\pder{S}{E})^{-1}$. 
  
  Systems for which a description based on 
  canonical ensemble is possible, the Laplace transform of $g(E)$ with respect to a variable
  $\beta$ will give the partition function $Z(\beta)$. It is, however, trivial to show that 
  gravitating systems of interest in astrophysics cannot be described by a canonical
  ensemble \cite{tppr}, \cite{dlbone}, \cite{dlbtwo}.
  Virial theorem holds for such systems  and we have $(2K+U) =0$ where 
  $K$ and $U$ are the total kinetic and potential energies of the system.
  This leads to $E=K+U= -K$; since the temperature of the system is proportional
  to the total kinetic energy, the  specific heat\index{specific heat} will be negative: 
  $C_V \equiv (\pder{E}{T})_V 
  \propto (\pder{E}{K}) < 0$.   On the other hand, the specific heat of any system
  described by a canonical ensemble $C_V = \beta^2 <(\Delta E)^2>$ will be
  positive definite. Thus one cannot describe self gravitating systems
  of the kind we are interested in by  canonical ensemble\index{canonical ensemble}.
  
  One may attempt to find the equilibrium configuration for self gravitating
  systems by maximizing the entropy $S(E)$ or the phase volume 
  $g(E)$. It is again easy to show that no global maximum for the entropy
  exists for classical point particles interacting via Newtonian gravity.
  To prove this, we only need to construct a configuration with   arbitrarily high
   entropy which can be achieved as follows:
  Consider a system of $N$ particles initially occupying a region of finite 
  volume in phase space and total energy $E$. We now move  a small
  number of these particles (in fact, a pair of them, say, particles 1 and 2  will do) arbitrarily close to 
  each other. The potential energy of interaction of  these two particles, 
  $-Gm_1m_2/r_{12}$, will become arbitrarily high as $r_{12}\to 0$. Transferring some
  of this energy to the rest of the particles, we can increase their kinetic energy without limit.
  This will clearly increase the phase volume occupied by the system without bound.
  This argument can be made more formal by dividing the original system into a small,
  compact core and a large diffuse halo and allowing the core to collapse
  and transfer the energy to   the halo.

  The absence of the global maximum for entropy --- as 
  argued above --- depends on the idealization
   that there is no short distance cut-off
  in the interaction of the particles, so that we could take the limit $r_{12}\to 0$.
  If we assume, instead, that each particle has a minimum radius $a$, then
  the typical lower bound to the gravitational potential energy contributed by a 
  pair of particles will be $-Gm_1m_2/2a$. This will put an upper bound on the 
  amount of energy that can be made available to the rest of the system.
  
   We have also assumed that part of the system can expand without limit --- in the sense
  that any particle with sufficiently large energy can move to arbitrarily large
  distances. In real life, no system is completely isolated and eventually one has to 
  assume that the meandering particle is better treated as a member of 
  another system. One way of obtaining a truly isolated system is to confine
  the system  inside a spherical region of radius $R$ with, say, reflecting
  wall. 
  
 The two cut-offs $a$ and $R$ will make the upper bound on the entropy finite, but
 even with  the two cut-offs, the primary nature
  of gravitational instability cannot be avoided. The basic phenomenon
  described  above  (namely, the formation of a  compact
  core\index{compact
  core} and a  diffuse halo\index{diffuse halo}) will still occur since this is the direction of increasing
  entropy. Particles in the hot diffuse component will permeate the entire spherical
  cavity, bouncing off the walls and having a kinetic energy which is 
  significantly larger than the potential energy. The compact core will exist as 
  a gravitationally bound system with very little kinetic energy. 
  A more formal way of understanding this phenomena is based 
  on  the virial theorem for a system with a short distance cut-off
  confined to a sphere of volume $V$. In this case, the virial theorem will
  read as \cite{textone} 
  \begin{equation}
  2T + U = 3PV + \Phi
  \label{modvirial}
  \end{equation}
  where $P$ is the pressure on the walls and $\Phi$ is the correction to the potential
  energy arising from the short distance cut-off. This equation can be satisfied
  in essentially three different ways. If $T$ and $U$ are significantly higher than
  $3PV$, then we have $2T + U \approx 0$ which describes a self gravitating 
  systems in standard virial equilibrium but not in the state of maximum entropy.
  If $T \gg U$ and $3PV\gg \Phi$, one can have $2T \approx 3PV$ which
  describes an ideal gas with no potential energy confined to a container of
  volume $V$; this will describe the hot diffuse component at late times.
  If  $T \ll U$ and $3PV\ll \Phi$, then one can have $U\approx \Phi$ describing
  the compact potential energy dominated core at late times.
  In general, the evolution of the system will lead to the production of the core and
  the halo and each component will satisfy the virial theorem in the form (\ref{modvirial}).
  Such an asymptotic state with two distinct phases  \cite{aaron}  is quite different from what would have
  been expected for systems with only short range interaction.
  Considering its importance, I shall briefly describe in section \ref{sec:phases} 
   a toy model which captures the essential
  physics of the above system in an exactly solvable context. 
  
  The above discussion focussed on the existence of global maximum
  to the entropy and we proved that it does not exist in the absence of 
  two cut-offs. It is, however, possible to have {\it local} extrema of entropy 
  which are not global maxima. Intuitively, one would have expected
  the distribution of matter in  the configuration which is a local
  extrema of entropy to be described by a Boltzmann distribution,
  with the density given by $\rho ({\bf x}) \propto \exp[-\beta \phi({\bf x})]$
  where $\phi$ is the gravitational potential related to $\rho$ by Poisson equation.
  This is indeed true and a formal proof will be given in section \ref{mfgrav}.  This
  configuration is usually called the  isothermal sphere\index{isothermal sphere} (because it can be 
  shown that, among all solutions to this equation, the one with spherical symmetry
  maximizes the entropy) and since it is a local maximum of entropy, it deserves
  careful study. I will describe briefly some of the interesting features of the isothermal
  spheres in section \ref{isosph}   and  this configuration will
  play a dominant role throughout our review. The second (functional)  derivative
  of the entropy with respect to the configuration variables will determine whether
  the local extremum of entropy is a local maximum or a saddle point \cite{antonov}, \cite{tpapjs}
   and
  some of these  results are described at the end of section \ref{isosph}. 
  The relevance of the long range of gravity in all the above phenomena can be
   understood by studying model
  systems with an attractive potential varying as $r^{-\alpha}$ with different values
  for $\alpha$. Such studies confirm the results and interpretation given above;
  (see \cite{ispo} and references cited therein).
  
  Let us now consider the situation in the context of an expanding background
  described in Part II
   which is the main theme of the review.
  There is considerable amount of observational evidence to suggest that
  one of the dominant energy densities in the universe is contributed by self gravitating
  point particles. The smooth average energy density of these particles drive
  the expansion of the universe while any small deviation from the homogeneous energy
  density will cluster gravitationally. One of the central problems in cosmology is to describe
  the non linear phases of this gravitational clustering starting from a initial spectrum of density
  fluctuations. It is often enough (and necessary) to use a statistical description and  relate
  different statistical indicators (like the  power spectra\index{power spectra}, $n$th order  correlation functions\index{correlation functions} ....)
  of the resulting density distribution to the statistical parameters (usually the power spectrum) of the 
  initial distribution. The relevant scales at which gravitational clustering is non linear are less than
  about 10 Mpc (where 1 Mpc = $3\times 10^{24}$ cm is the typical separation between galaxies in the
  universe) while the expansion of the universe has a characteristic scale of about few thousand 
  Mpc. Hence, non linear gravitational clustering in an expanding universe can 
  be adequately described by Newtonian gravity provided the rescaling of lengths due to the
  background expansion
  is taken into account. This is easily done by introducing a  {\it proper}
   coordinate\index{proper
   coordinate} for the $i-$th particle ${\bf r}_i$,
  related to the  {\it comoving} coordinate\index{comoving coordinate}  ${\bf x}_i$, by ${\bf r}_i = a(t) {\bf x}_i$ with 
  $a(t)$ describing the stretching of length scales due to cosmic expansion. The Newtonian 
  dynamics works with the proper coordinates ${\bf r}_i$ which can be translated
  to the behaviour of the comoving coordinate ${\bf x}_i$ by this rescaling.
  (Some basic results in cosmology are summarized in Appendix A.)
  
  As to be expected, cosmological expansion  completely changes the nature of the problem because of 
  several new factors which come in:
  (a) The problem has now become time dependent and it will be pointless to look for
  equilibrium solutions in the conventional sense of the word. 
  (b) On the other hand, the expansion of the universe has a civilizing influence on the 
  particles and  acts counter to the tendency of gravity
  to make systems unstable. 
  (c) In any small local region of the universe, one would assume that the conclusions
  describing a finite gravitating system will still hold true approximately. In that case, particles in any small
  sub region will be driven towards configurations of local extrema of entropy (say, isothermal
  spheres) and towards global maxima of entropy (say, core-halo configurations).
  
  An extra feature comes into play as regards the expanding halo from any sub region.
  The expansion of the universe acts as a damping term in the equations of motion
  and drains the particles of their kinetic energy --- which is essentially the lowering of
  temperature of any system participating in cosmic expansion.  This, in turn, 
  helps gravitational clustering since the potential wells of nearby sub regions
  can capture particles in the expanding halo of one region when the kinetic energy of 
  the expanding halo  has been sufficiently reduced. 
  
  The actual behaviour of the system will, of course, depend on the form of $a(t)$.
  However, for understanding the nature of clustering, one can take $a(t) \propto 
  t^{2/3}$ which describes a matter dominated universe with critical density (see Appendix A).
  Such a power law has the advantage that there is no intrinsic scale in the problem.
  Since Newtonian gravitational force is also scale free, one would expect some 
  scaling relations to exist in the pattern of gravitational clustering. Incredibly enough,
  converting this intuitive idea into a concrete mathematical statement
  turns out to be non trivial and difficult. I shall discuss several attempts
  to give concrete shape to this idea in sections \ref{renormgrav}, \ref{nlscales} and \ref{univgravcl}
     but
  there is definite scope for further work in this direction.

 To make any progress we need a theoretical
  formulation which will  relate statistical indicators in the non linear regime of 
  clustering to the initial conditions. In particular, we need
  a robust prescription which will allow us to obtain the two-point correlation function 
  and the nonlinear  power spectrum from the initial power spectrum. Fortunately, this problem
  has been solved to a large extent and hence one can use this as a basis for attacking
  several other key questions.
  There are four key theoretical questions which are of considerable interest in this area
  that I will focus on:

\begin{itemize}
\item
   If the initial power spectrum is sharply peaked in a narrow band of wavelengths, how does
  the evolution transfer the power to other scales? This is, in some sense, analogous to determining
  the  Green function for the gravitational clustering\index{Green function for the gravitational clustering} except that superposition will not work
  in the non linear context.  
  \item
   What is the asymptotic nature of evolution for the self gravitating system in an expanding
  background? In particular, how can one connect up the local behaviour of gravitating systems
  to the overall evolution of clustering in the universe?
  (If we assume that the isothermal spheres play an important role in the local
  description of gravitating system, we would expect a strong trace of it to survive
  even in the context of cosmological clustering. This is indeed true as I shall show
  in sections \ref{nlscales}, \ref{cipt} and \ref{univgravcl}  
  but only in the asymptotic limit, under certain assumptions.) 
 \item
  Does the gravitational clustering at late stages wipe out the memory of initial conditions or does
  the late stage evolution depend on the initial power spectrum of fluctuations?
\item
   Do the  virialized structures\index{virialized structures} formed in an expanding universe due to gravitational
  clustering have any invariant properties? Can their structure be understood from first principles?
\end{itemize}

  All the above questions are, in some sense, open and thus are good research problems.
  I will highlight the progress which has been made and give references to original 
  literature for more detailed discussion.
  
  \vskip 1cm
  
 \noindent {\Large{\bf Part I: Gravitational clustering in static backgrounds  }}
 
 \vskip 0.2cm

\section{\label{sec:phases}Phases of the self gravitating system}

As described in section \ref{intro}  the statistical mechanics of
finite, self gravitating,
systems have the following characteristic features: 
(a) They exhibit negative specific heat while in virial equilibrium.
(b) They are inherently unstable to the formation of a core-halo structure and global maximum for entropy
does not  exist without cut-offs at short and large distances. 
(c) They can be broadly characterized by two phases --- one of which is compact and dominated
by potential energy while the other is diffuse and behaves more or less like an ideal gas.
The purpose of this section is to describe a simple toy model which exhibits all these 
features and mimics a self gravitating system \cite{tppr}.  

Consider a system with two particles described by a Hamiltonian of the form
\begin{equation}H \left(  {\bf P}, {\bf Q}; {\bf p}, {\bf r}\right) = {{\bf P}^2 \over 2M} + {{\bf p}^2 \over 2 \mu} - {Gm^2 \over r }   \end{equation}
where $(  {\bf Q}, {\bf P}) $ are coordinates and momenta of the center of mass, $(  {\bf r}, {\bf p} )$ are the  relative coordinates and momenta, $M = 2 m$ is the total mass, $\mu = m/2 $ is the reduced mass  and $m$ is the mass of the individual particles. This system may be thought of as consisting of  two particles (each  of mass $m$) interacting via gravity. We shall  assume that the quantity $r$ varies in the interval $(  a, R )$. This is equivalent to assuming that the particles are hard spheres of radius $a/2$ and that the system is confined to a spherical box of radius $R$. We will study the ``statistical mechanics'' of this  simple toy model\index{simple toy model}.

 To do this, we shall start with the volume $g(E)$
of the constant energy surface $H=E$.
Straightforward calculation gives
\begin{equation}
 g(E)=  AR^3\int_a^{r_{\rm max}} r^2 dr \left[E+{Gm^2\over r}\right]^2.\label{pv}
 \end{equation} 
where $A=(64 \pi ^5 m^3/3)$.
The range of integration in (\ref{pv})  should be limited to the region
in which the expression in the square brackets is positive. So we
should use $r_{\rm max}= (Gm^2/|E|)$ if $ (-Gm^2/a)< E < (-Gm^2/R) $, and use
$r_{\rm max}= R$ if $(-Gm^2/R)<E<+\infty$. Since $H\geq (-Gm^2/a)$, we trivially
have $g(E)=0$ for $E<(-Gm^2/a)$. The constant $A$ is unimportant for our
discussions and hence will be omitted from the formulas hereafter.
The integration in (\ref{pv})  gives the following result:
\begin{equation}
 {g(E)\over (Gm^2)^3}=\left\{ 
 \begin{array}{l}
  {R^3\over3}(-E)^{-1} \left( 1+{aE\over Gm^2}\right)^3 ,\quad\qquad
           (-Gm^2/a)< E<(-Gm^2/R)\\
           {   }\\
          {R^3\over3} (-E)^{-1} \left[ \left(1+{RE\over Gm^2}\right)^3
          - \left(1+{aE\over Gm^2}\right)^3 \right],
         (-Gm^2/R )<E< \infty .
 \end{array}
        \right.
         \label{gee}\end{equation}
 This function $g(E)$ is continuous and smooth at $E= (-Gm^2/R)$. We define the entropy $S(E)$  and
the temperature $T(E)$ of
the system by the relations  
\begin{equation} S(E)=\ln g(E);\quad T^{-1}(E)=\beta (E)= {\partial S(E)\over \partial E}.
\label{entro}\end{equation}        
All the interesting thermodynamic properties of the system can be understood
from the $T(E)$ curve.

Consider first the case of  low energies with $(-Gm^2/a)<E<(-Gm^2/R)$.
Using (\ref{gee}) and (\ref{entro})  one can easily obtain $T(E)$ and write it in the
 dimensionless form as
\begin{equation} t(\epsilon)= \left[{3\over 1+\epsilon}
 -{1\over \epsilon}\right]^{-1}\label{temp}\end{equation}
where we have defined $t=(aT/Gm^2)$ and $\epsilon =(aE/Gm^2)$.

This function exhibits the peculiarities characteristic of gravitating systems.
At the lowest energy admissible for our system, which corresponds to $\epsilon
=-1$, the temperature $t$ vanishes. This describes a tightly bound
low temperature phase of the system with negligible random motion. 
 The $t(\epsilon)$
is clearly dominated by the first term of (\ref{temp})   for $\epsilon \simeq -1$. As we increase
the energy of the system, the temperature {\it increases}, which is the normal
behaviour for a system. This trend continues up to
\begin{equation}\epsilon =
\epsilon_1= -{1\over2}(\sqrt3 -1) \simeq -0.36  \end{equation}
at which point the $t(\epsilon)$ curve reaches a maximum and turns around.
As we increase the energy further the temperature {\it decreases}.  The
system {\it exhibits negative specific heat in this range.}

Equation (\ref{temp})  is valid  from the minimum energy $(-Gm^2/a)$
 all the way up to the energy $(-Gm^2/R)$. For realistic systems, $R\gg a$
and hence this range is quite wide. For a small region in this range,
[from $(-Gm^2/a)$ to $(- 0.36 Gm^2/a)$] we have positive specific heat;
for the rest of the region the specific heat is negative. {\it The positive
specific heat region owes its existence to the nonzero short distance
cutoff.} If we set $a=0$, the first term in (\ref{temp})   will vanish; we will
have $t\propto (-\epsilon^{-1})$ and negative specific heat in this entire domain.

For $E\geq (-Gm^2/R)$, we have to use the second expression in ({\ref{gee})  for $g(E)$.
In this case, we get:
\begin{equation} t(\epsilon)= \left[ { 3\left[ (1+\epsilon)^2- {R\over a}(1+ {R\over a}\epsilon)^2\right]
                        \over (1+\epsilon)^3 -(1+{R\over a}\epsilon)^3}
                       -{1\over \epsilon} \right]^{-1} .\label{qtemp}\end{equation}
This function, of course, matches smoothly with (\ref{temp})   at $\epsilon=-(a/R)$.
As we increase the energy, the temperature continues to decrease for
a little while, exhibiting negative specific heat. However, this behaviour
is soon halted at some $\epsilon=\epsilon_2$, say. The $t(\epsilon)$ curve reaches
a minimum at this point, turns around, and starts increasing with increasing
$\epsilon$. We thus enter another (high-temperature) phase with positive
specific heat. From (\ref{qtemp})   it is clear that $t\simeq (1/2)
 \epsilon$ for large $\epsilon$. 
(Since $E=(3/2)NkT$ for an ideal gas, we might have expected to find
$t\simeq (1/3)\epsilon$ for our system with $N=2$ at high temperatures.
This is indeed what we would have found if we had defined our entropy as
the logarithm of the volume of the phase space with
$H \le E$. With our definition, the energy of the ideal gas is actually
$E=[{(3/2)}N-1]kT;$ hence we get $t=(1/2)\epsilon$ when $N=2$). 
The form of the $t(\epsilon)$ for $(a/R) = 10^{-4}$  is shown
in figure \ref{fig:phases} by the dashed curve.  The specific heat is positive along the portions AB and CD
and is negative along BC.

%%%%%%%%%%%%%%%%%%%%%%%%%%%%%%%%%%%%%%% 
     \begin{figure}[ht]
   \begin{center}
   \includegraphics[width=0.85\textwidth]{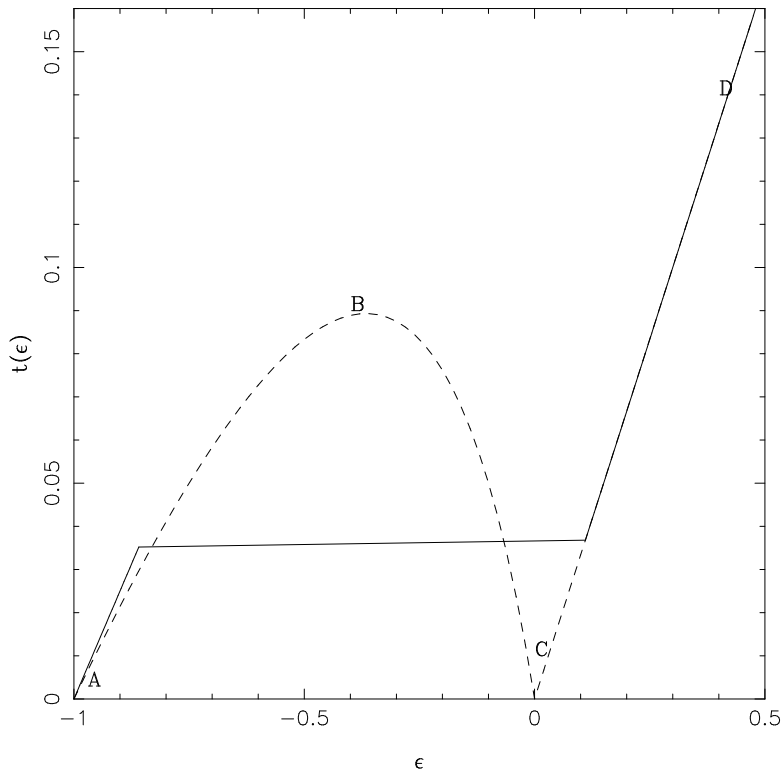}
   \end{center}
   \caption{The relation between temperature and energy for a model mimicking self
   gravitating systems. The dashed line is the result for micro-canonical
   ensemble and the solid line is for canonical ensemble. 
   The negative specific heat region, BC, in the micro-canonical 
   description is replaced by a phase transition in the canonical
   description. See text for more details   }
   \label{fig:phases}
  \end{figure}
%%%%%%%%%%%%%%%%%%%%%%%%%%%%%%%%%%%%%%

The overall picture is now clear. Our system has two natural energy scales:
$E_1=(-Gm^2/a)$ and $E_2=(-Gm^2/R)$. For $E\gg E_2$, gravity is not strong
enough to keep $r<R$ and the system behaves like a gas confined
by the container; we have a  high temperature phase\index{high temperature phase} with positive specific
heat. As we lower the energy to $E\simeq E_2$,
 the effects of gravity begin to be felt.
For $ E_1<E<E_2$, the system is unaffected by either the box or the short
distance cutoff; this is the domain dominated entirely by gravity and
we have negative specific heat. As we go to $E\simeq E_1$,
the hard core nature of the particles begins to be felt and the gravity is
again resisted. This gives rise to a  low temperature phase\index{low temperature phase} with positive
specific heat.

 We can also consider the effect of increasing $R$, keeping
$a$ and $E$ fixed. 
Since we imagine the particles to be hard spheres of radius $(a/2)$,
we should only consider $R>2a$. It is amusing to note that, if
$2<(R/a)<(\sqrt 3 +1)$, there is no region of negative specific heat.
As we increase $R$, this negative specific heat region appears
 and increasing $R$  increases
the range over which the specific heat is negative. Suppose a
system is originally prepared with some $E$ and $R$ values such that the specific heat 
is positive. If we now increase $R$,  the system may find itself
in a region of negative specific heat.{\it This suggests the possibility that
an instability may be triggered in a constant energy system if its radius
increases beyond a critical value.}  We will see later that this is indeed true.

Since systems described
by canonical distribution cannot 
exhibit negative specific heat, it follows that canonical distribution will
lead to a very different physical picture for this range of (mean) energies $E_1<E<E_2$.
It is, therefore, of interest to look at our system from the point of view of
canonical distribution by computing the partition function.
In the partition function
\begin{equation} Z(\beta)=\int d^3Pd^3pd^3Qd^3r \exp (-\beta H)  \end{equation}
 the integrations over $P,p$ and $Q$ can be performed trivially.
Omitting an overall constant which is unimportant, we can write the
answer  in the dimensionless form as
\begin{equation} Z(t)= t^{3} \left({R\over a}\right)^3\int_1^{R/a} dx
 x^2 \exp \left( {1\over xt}\right) \label{parfn}\end{equation}
where $t$ is the dimensionless temperature defined in (\ref{temp})  . Though this
integral cannot be evaluated in closed form, all the limiting properties of
$Z(\beta)$ can be easily obtained from (\ref{parfn}). 

The integrand in (\ref{parfn})  is large for both large and small $x$ and reaches
a minimum for $x=x_m=(1/2t)$.  At high temperatures, $x_m <1$ and hence the minimum falls outside
the domain of integration. The exponential contributes very little to
the integral and we can approximate $Z$ adequately by
\begin{equation}Z\approx t^3\left({R\over a}\right)^3
\int_1^{R/a} dx x^2 \left[1+{2x_{m}\over x}\right]={t^3\over 3}
\left({R\over a}\right)^6\left(1+{3a\over 2Rt}\right)
.\label{hparfn}\end{equation}
On the other hand, if $x_m>1$ the minimum lies between the limits of the
integration and the exponential part of the curve dominates the integral.
We can easily evaluate this contribution by a saddle point approach, and obtain 
\begin{equation}
 Z \approx \left({R\over a}\right)^3t^4 (1-2t)^{-1} \exp\left( {1\over t}\right)  .
\label{lparfn}\end{equation}
As we lower the temperature, making $x_m$ cross $1$ from below, the contribution
switches over from (\ref{hparfn})  to (\ref{lparfn}). The transition is exponentially sharp. The  critical
temperature\index{critical
temperature} at which the transition occurs can be estimated by finding the temperature
at which the two contributions are equal. This occurs at
\begin{equation} t_c={1\over 3}{1\over \ln (R/a)}.\label{tcrit}\end{equation}
For $t<t_c$, we should use (\ref{lparfn})  and for $t>t_c$ we should use (\ref{hparfn}).

Given $Z(\beta)$ all thermodynamic functions can be computed. In particular,
the mean energy of the system is given by
$ E(\beta)= -(\partial \ln Z/ \partial\beta)$.
This relation can be inverted to give the $T(E)$ which can be compared
with the $T(E)$ obtained earlier using the micro-canonical distribution.
 From (\ref{hparfn})  and (\ref{lparfn})  we get,
\begin{equation}\epsilon (t)= {aE\over Gm^2}= 4t-1  \label{lener}\end{equation}
 for $t<t_c$ and
\begin{equation}\epsilon(t)= 3t-{3a\over 2R} \label{hener}\end{equation}
for $t>t_c$. Near $t\approx t_c$, there is a rapid variation of the energy
and we cannot use either asymptotic form. The system undergoes a phase transition
at $t=t_c$ absorbing a large amount of energy
\begin{equation}\Delta \epsilon\approx \left( 1-{1\over 3\ln (R/a)}\right). \end{equation}
The specific heat is, of course, positive throughout the range. This is
to be expected because canonical ensemble cannot lead to negative specific
heats.

The  $T-E$ curves obtained from the canonical (unbroken line)  and micro-canonical (dashed line)
distributions are shown in figure \ref{fig:phases}. 
 (For convenience, we have rescaled the $T-E$ curve of the micro-canonical distribution so that $\epsilon \simeq 3t$ asymptotically.)
At both very low and very high temperatures,   the  canonical and micro-canonical 
descriptions
match. The crucial difference occurs at the intermediate energies and
temperatures. Micro-canonical description predicts negative specific heat and
a reasonably slow variation of energy with temperature. Canonical
description, on the other hand, predicts a phase transition with
rapid variation of energy with temperature. Such phase transitions are
accompanied by large fluctuations in the energy, which is the main reason
for the disagreement between the two descriptions \cite{tppr}, \cite{dlbone}, \cite{dlbtwo}.  

Numerical analysis of more realistic systems confirm all these features. 
Such systems exhibit a phase transition from the  diffuse virialized phase\index{diffuse virialized phase}
to a  core dominated phase\index{core dominated phase} when the temperature is lowered below a critical value \cite{aaron}.
The transition is very sharp and occurs at nearly constant temperature.
The energy released by the formation of the compact core heats up the diffuse
halo component.

 \section{\label{mfgrav}Mean field description of gravitating systems}

   The analysis in the previous two sections  shows that there is no {\it global} maximum
   for the entropy for a self gravitating system of point particles and the evolution will proceed
   towards the formation of a core halo configuration and will continue inexorably
   in the absence of cut-offs. It is, however, possible to find configurations for these 
   systems which are {\it local} maxima of the entropy. 
   This configuration, called the isothermal sphere, will be of considerable 
   significance in our discussions.

Consider a system of $N$ particles interacting with each other through
the two-body potential $U({\bf x}, {\bf y})$. The entropy $S$ of this 
system, in the micro-canonical description, is defined through the relation
\begin{equation} {\rm e}^S =g(E)= {1\over N!}\int d^{3N}x d^{3N}p \delta (E-H)
                ={A\over N!} \int d^{3N}x \left[E- {1\over2}\sum _{i\not= j}
                 U({\bf x}_i,{\bf x}_j)\right]^{3N\over 2}    \label{qgee}\end{equation}
wherein we have performed the momentum integrations and replaced $(3N/2 -1)$
by $(3N/2)$. We shall approximate the expression in (\ref{qgee})  in the following manner.

Let the spatial volume $V$ be divided into $M$ (with $M\ll N$) cells of equal size, large enough
to contain many particles but small enough for the potential to be treated as a constant 
inside each cell. 
Instead of integrating over the particle coordinates $({\bf x}_1,{\bf x}_2,...,{\bf x}_N)$
we shall sum over the number of particles $n_a$ in the cell centered at ${\bf x}_a$
(where $a=1,2,...,M$). Using the standard result that the integration over $(N!)^{-1}d^{3N}x$
 can be replaced by 
\begin{equation} \sum_{n_1=1}^\infty \left( {1\over n_1!}\right)  \sum_{n_2=1}^\infty \left( {1\over n_2!}\right) ....
 \sum_{n_M=1}^\infty \left( {1\over n_M!}\right) \;
\delta \left( {N-\sum_a n_a}\right)  \; \left(V\over M \right)^N  \end{equation}
and ignoring the unimportant constant $A$, we can rewrite (\ref{qgee})  as
\begin{eqnarray} {\rm e}^S & =& \sum_{n_1=1}^\infty \left( {1\over n_1!}\right) 
 \sum_{n_2=1}^\infty \left( {1\over n_2!}\right) ..
 \sum_{n_M=1}^\infty \left( {1\over n_M!}\right) 
\delta \left( {N-\sum_a n_a}\right)   \left( {V\over M}\right) ^N\nonumber\\
&&\hskip8em \times \qquad \left[E- {1\over2}
 \sum _{a\not= b}^{M} n_a U_{ab} n_b \right]^{3N\over 2}\nonumber\\
&\approx& \sum_{n_1=1}^\infty  \sum_{n_2=1}^\infty ....
 \sum_{n_M=1}^\infty \;\delta \left( {N-\sum_a n_a}\right)  \;\exp S[\{n_a\}]
  \label{qsum}\end{eqnarray}
where
\begin{equation} S[\{n_a\}]= {3N\over 2} \ln \left[ E- {1\over2}\sum_{a\not= b}^M
n_a U({\bf x}_a,{\bf x}_b) n_b \right]
- \sum_{a=1}^M n_a \ln \left( {n_aM\over {\rm e} V}\right) .\label{qentro}\end{equation}
In arriving at the last expression we have used the Sterling's
 approximation for the factorials. The mean 
field limit is now obtained by retaining in the sum in (\ref{qsum})  only the term for which
the summand reaches the maximum value, subject to the constraint on the total
number. That is, we use the approximation
\begin{equation} \sum_{\{ n_a \} } {\rm e}^{S[n_a]} \approx {\rm e}^{S[n_{a,{\rm max}}]}\label{qapsum}\end{equation}
where $ n_{a,{\rm max}}$ is the solution to the variational problem 
\begin{equation} \left({\delta S\over \delta n_a} \right)_{n_a=n_{a,{\rm max}}}=0 \quad\quad
 {\rm with}\quad \sum_{a=1}^M n_a= N . \end{equation}
Imposing this constraint by a Lagrange multiplier and using the expression (\ref{qentro})  
for $S$, we obtain the equation satisfied by $n_{a,{\rm max}}$:
\begin{equation} {1\over T} \sum_{b=1}^M U({\bf x}_a,{\bf x}_b)n_{b,{\rm max}}
 + \ln \left( {n_{a,{\rm max}} M \over V}\right) = {\rm constant}\label{qextr}\end{equation}
where we have defined the temperature $T$ as
\begin{equation} {1\over T}=  {3N\over 2}\left (
E-{1\over2}\sum_{a\not= b}^M n_a U({\bf x}_a,{\bf x}_b)n_b\right)^{-1}=\beta .\label{eqtemp}\end{equation}
We see from (\ref{qentro})  that this expression is also equal to $({\partial S/\partial E})$;
 therefore $T$ is indeed the correct thermodynamic temperature. 
 We can now return back to the continuum limit by the replacements
\begin{equation} n_{a,{\rm max}}{M\over V}=\rho ({\bf x}_a);\quad\quad
 \sum_{a=1}^M \rightarrow {M\over V} \int . \end{equation}
In this limit, the extremum solution (\ref{qextr})  is given by  
\begin{equation} \rho ({\bf x})= A \exp (-\beta \phi ({\bf x}));\quad {\rm where }
  \quad\phi({\bf x})= \int d^3{\bf y} U({\bf x},{\bf y}) \rho ({\bf y}) \end{equation}
which, in the case of gravitational interactions, becomes
\begin{equation}\rho({\bf x})= A \exp(-{\beta\phi({\bf x})}); \quad 
\phi({\bf x})=-G\int{\rho({\bf y})d^3{\bf y}\over
|{\bf x}-{\bf y}|}.\label{qiso}\end{equation}
Equation (\ref{qiso})  represents the  configuration of extremal entropy for a gravitating
system in the mean field limit. The constant $\beta$ is already determined
through (\ref{eqtemp})   in terms of the total energy of the system. The constant $A$ has to
be fixed in terms of the total number (or mass) of the particles in the system.

 The various manipulations in (\ref{qgee})  to (\ref{qextr})  tacitly assume that 
 the expressions we are dealing  with
are finite. But for gravitational interactions {\it without} a short distance
cut-off, the quantity ${\rm e}^S$ - and hence all the terms we have been handling - 
 are divergent. We should, therefore, remember that a short distance cut-off is
needed to justify the entire procedure \cite{fola}, \cite{ispo}, \cite{horkatz} and that (\ref{qiso}) --- which is based on
a strict $r^{-1}$ potential and does not
incorporate any such cutoff --- can only be approximately correct.
We will now study some of the properties of the solution of (\ref{qiso})  which is a local extremum
of the entropy.

\section{\label{isosph}Isothermal sphere}

The extremum condition for the entropy,
equation (\ref{qiso}), is equivalent to the 
  differential equation
for the gravitational potential:
\begin{equation}\nabla^2 \phi = 4 \pi G \rho_c e^{-{\beta} \left[ \phi \left(  {\bf x} \right)  - \phi \left(  0 \right) 
 \right] } \label{qself} \end{equation}
Given the solution to this equation, all other quantities can be
determined. As we shall see, this system shows
several peculiarities.

It is convenient to introduce the length, mass and energy scale by the definitions
\begin{equation}L_0 \equiv \left(  4 \pi G \rho_c \beta \right) ^{1/2}, \quad M_0 = 4 \pi \rho_c L_0^3, \quad\phi_0 \equiv \beta^{-1} = {GM_0 \over L_0 }   \end{equation}
where $\rho_c = \rho(0)$. All
 other physical variables can be expressed in terms of the 
dimensionless quantities
\begin{equation} x \equiv {r \over L_0}, \quad n \equiv {\rho \over \rho_c }, \quad m= {M \left(  r \right)  \over M_0 }, \quad y \equiv \beta \left[ \phi - \phi \left(  0 \right)  \right] .  \end{equation}
In terms of $y(x)$ the isothermal equation (\ref{qself}) becomes 
\begin{equation}{1\over x^2}{d\over dx}(x^2{dy\over dx})={\rm e}^{-y}\label{qdliso}\end{equation}
with the boundary condition $y(0)=y'(0)=0$. Let us consider the nature of
solutions to this equation.

By direct substitution, we see that $n = \left(  2 /x^2 \right) , m = 2x, y = 2 \ln x $ satisfies these equations. This solution, however, is singular at the origin
and hence is not physically admissible. The importance of this solution
lies in the fact that  other (physically admissible) solutions
tend to this solution \cite{tppr}, \cite{chandra} for large values of $x$. 
This asymptotic behavior of all solutions shows that the density decreases as $(1/r^2)$ for
large $r$ implying that the mass contained inside a sphere of
radius $r$ increases as $M(r)\propto r$ at large $r$. To find physically 
useful solutions, it is necessary to assume that the solution is cutoff at
some radius $R$. For example, one may assume that the system is enclosed
in a spherical box of radius $R$.  In what follows, it will be assumed that
the system has some cutoff radius $R$.

 The equation  (\ref{qdliso})  is invariant under the transformation $y\rightarrow y+a \; ; \;
x\rightarrow kx $ with $k^2= {\rm e}^a$. This invariance implies that,
given a solution with some value of $y(0)$, we can obtain the solution
with any other value of $y(0)$ by simple rescaling. Therefore, only one
of the two integration constants in (\ref{qdliso}) is really non-trivial. 
Hence it must be possible  to reduce the degree of the equation from two to one
by a judicious choice of variables \cite{chandra}. 
One such set of variables are:
\begin{equation}v\equiv{m\over x};\quad u\equiv{nx^3\over m}={nx^2\over v}. \end{equation}
In terms of $v$ and  $u$, equation (\ref{qself}) becomes
\begin{equation}{u\over v}{dv\over du}=-{(u-1)\over (u+v-3)}.\label{quv}\end{equation}
The boundary conditions $y(0) = y'(0)=0$ translate into the
following: $v$ is zero at $u=3$, and $(dv/du)=-5/3$ at (3,0). The solution $v\left(  u \right) $ has to be obtained numerically: it is plotted in figure \ref{figisotherm} 
as the spiraling curve. The singular points of this differential equation are
given by the intersection of the straight lines $u=1$ and $u+v=3$ on which, the
numerator and denominator of the right hand side of (\ref{quv}) vanishes; that is,
the singular point is at $u_s =1$, $ v_s =2$ corresponding to the solution $n = (2/x^2), m=2x$. It is obvious from the 
nature of the equations that the solutions will spiral around the singular
point.

 The nature of the solution shown in figure \ref{figisotherm}  allows us
to put an interesting bounds on physical quantities
including energy.  To see this,
we shall compute the total energy $E$ of the isothermal sphere. The potential
and kinetic energies are
\begin{eqnarray}
U&=&-\int^R_0{GM(r)\over r} {dM\over dr}
dr = - {GM_0^2\over L_0} \int^{x_0}_0 mnxdx \nonumber\\ 
K&=&{3\over 2}{M\over \beta}={3\over 2}{GM^2_0\over L_0}
m(x_0) = {GM^2_0\over L_0}{3\over 2}\int^{x_0}_0
nx^2 dx \end{eqnarray}
where $x_0=R/L_0$. The total energy is, therefore, 
\begin{eqnarray}
E&=&K+U={GM^2_0\over 2L_0}\int^{x_0}_0 dx
(3nx^2-2mnx)\nonumber\\ 
 &=& {GM^2_0\over 2L_0}\int^{x_0}_0
dx{d\over dx}\{2nx^3-3m\}
= {GM^2_0\over L_0}\{n_0 x_0^3-{3\over 2}m_0\}  \end{eqnarray}
where $n_0=n(x_0)$ and $m_0=m(x_0)$. The dimensionless quantity
$(RE/GM^2)$ is given by
\begin{equation}\lambda={RE\over GM^2}={1\over v_0}
\{u_0-{3\over 2}\}. \end{equation}
{\it Note that the combination $(RE/GM^2)$ is a function of
$(u,v)$ alone}. Let us now consider the constraints on $\lambda$. 
Suppose we specify some value for $\lambda$ by specifying $R,E$ and
$M$. Then such an isothermal sphere {\it must} lie on the curve
\begin{equation}v={1\over \lambda}\left(  u-{3\over 2}\right) ; \qquad \lambda \equiv \frac{RE}{GM^2}\label{qeline}\end{equation}
which is a straight line through the point $(1.5,0)$ with the
slope $\lambda^{-1}$. On the other hand, since {\it all} isothermal spheres 
must lie on the $u-v$ curve, {\it an isothermal sphere can exist only
if the line in (\ref{qeline}) intersects the $u-v$ curve}.

\begin{figure}[ht]
\begin{center}
\includegraphics[width=.8\textwidth]{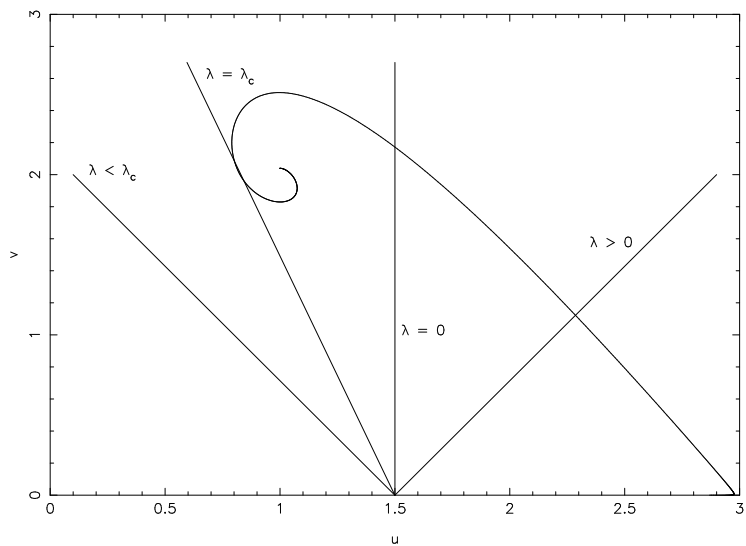}
\end{center}
\caption[]{Bound on $RE/GM^2$ for the isothermal sphere}
\label{figisotherm}
\end{figure}

 For large positive $\lambda$ (positive $E$)
there is just one intersection. When $\lambda=0$, (zero energy) we still have
a unique isothermal sphere. (For $\lambda =0 $, equation (\ref{qeline}) is a vertical line
through $u=3/2$.). When $\lambda$ is negative (negative $E$), the line can cut
the $u-v$ curve at more than one point; thus more than one isothermal 
sphere can exist with a given value of $\lambda$. [Of course, specifying 
$M,R,E$ individually will remove this non-uniqueness]. But as we decrease
$\lambda$ (more and more negative $E$) the line in (\ref{qeline}) will slope more
and more to the left; and when $\lambda$ is smaller than a critical value
$\lambda_c$, the intersection will cease to exist. {\it Thus no isothermal sphere
can exist if $(RE/GM^2)$ is below a  critical value $\lambda_c$.}\footnote{This derivation
is due to the author \cite{tpapjs}.
It is surprising that Chandrasekhar, who
has worked out the isothermal sphere in uv coordinates as early as 1939, missed discovering
the energy bound shown in figure \ref{figisotherm}.
 Chandrasekhar \cite{chandra} has the uv curve but does not 
over-plot lines of constant $\lambda$. If he had done that, he would have discovered  Antonov
instability\index{Antonov
instability} decades before Antonov did \cite{antonov}.}
This fact follows immediately from the nature of $u-v$ curve 
and equation (\ref{qeline}). The value of $\lambda_c$ can be found from the numerical solution in figure. It turns out to be about ($-0.335$). 

The isothermal sphere  has a special status as  a solution to the mean field
 equations. 
Isothermal spheres, however,  cannot exist if $(RE/GM^2) < -0.335$. Even when $(RE/GM^2)>-0.335$, the isothermal solution need not be stable. The stability of this solution can be investigated by studying the second variation of the entropy.
Such a detailed analysis shows that the following results are true \cite{antonov}, \cite{dlbwood},
\cite{tpapjs}.  
(i) Systems with $(RE/GM^2)<-0.335$ cannot evolve into isothermal
spheres. Entropy has no extremum for such systems.
(ii) Systems with ($(RE/GM^2)>-0.335$) and ($\rho(0)> 709\,\rho(R)$) can
exist in a meta-stable (saddle point state) isothermal sphere configuration. Here $\rho(0)$ and $\rho(R)$ denote the densities at the center and edge respectively. The entropy extrema exist but they are not local maxima.
(iii) Systems with ($(RE/GM^2)> -0.335$) and ($\rho(0)<709\,\rho(R)$) can
form isothermal spheres which are local maximum of entropy.

  \vskip 0.5cm
  
  \noindent {\Large {\bf Part II. Gravitational clustering in expanding universe}}
  
  \vskip 0.5cm

\noindent
Let us next consider the gravitational clustering of a system of collision-less point particles 
{\it in an expanding universe} which poses
several challenging theoretical questions. Though the problem can be
tackled in a `practical' manner using high resolution numerical simulations,
such an approach hides the physical principles which govern 
the behaviour of the system. To understand the physics, it is necessary that we
attack the problem from several directions using analytic and semi analytic
methods. These sections will describe such attempts and will emphasize
the semi analytic approach and outstanding issues, rather than more well established results. 
Some basic  results in cosmology are summarized in Appendix A.

\section{\label{gravclnl}Gravitational clustering at nonlinear scales}

The expansion of the universe sets a natural length scale (called the Hubble radius) $d_H = c
(\dot a/a)^{-1}$ which is about 4000 Mpc in the current universe. Since the non linear effects due
to gravitational clustering occur at significantly smaller length scales, it is possible to use
Newtonian gravity to describe these phenomena. 
In any region small compared to $d_{\rm H}$ one can set up an unambiguous coordinate system in which the {\it proper} coordinate of a particle ${\bf r} (t)=a(t){\bf x}(t)$ satisfies the Newtonian equation $\ddot {\bf r} = -  {\nabla }_{\bf r}\Phi$ where $\Phi$ is the gravitational potential. Expanding $\ddot \bld r$ and writing $\Phi = \Phi_{\rm FRW} + \phi$ where $\Phi_{\rm FRW}$ is due to the smooth (mean) density 
of matter  and $\phi$ is due to the perturbation in the density, we get
\begin{equation}
\ddot a {\bf x} + 2 \dot a \dot{\bf x} + a\ddot{\bf x} = - \nabla_{\bf r} \Phi_{\rm FRW} - \nabla_{\bf r}\phi = - \nabla_{\bf r} \Phi_{\rm FRW} - a^{-1} \nabla_{\bf x} \phi
\end{equation}
The first terms on both sides of the equation $\lb \ddot a{\bf x} \  {\rm and} -\nabla_{\bld r} \Phi_{\rm FRW} \rb$ should match since they refer to the global expansion of the background FRW universe
(see equation (\ref{cosacc}) of Appendix A). Equating them individually gives the results
\begin{equation} 
\ddot{\bf x} + 2 {\dot a \over a}\dot{\bf x} = - {1 \over a^2} \nabla_x \phi\ ; \qquad \Phi_{\rm FRW} = - {1 \over 2}{\ddot a \over a} r^2 = - {2\pi G \over 3}\rho_b r^2 
\end{equation}
where $\phi$ is the gravitational potential generated by the perturbed, Newtonian, mass density through 
\begin{equation}
 \nabla^2_x \phi = 4 \pi Ga^2(\delta \rho) = 4 \pi G \rho_ba^2 \delta . \end{equation}
If ${\bf x}_i(t)$ is the trajectory of the $i-$th particle, then equations for 
 gravitational clustering in an expanding universe, in the Newtonian
 limit, can be summarized as 
\begin{equation}
\ddot{\bf x}_i + { 2\dot a \over a} \dot{\bf x}_i = - {1 \over a^2} \nabla_{\bf x}
\phi;\quad \nabla_x^2 \phi = 4\pi G a^2 \rho_b \delta  \label{twnine}
\end{equation}
where $\rho_b(t)$ is the smooth background density of matter. We stress that, in the non-relativistic limit,
 the perturbed potential $\phi$ satisfies the usual Poisson equation.

Usually one is interested in the evolution of the density contrast $\delta \lb t, \bld x \rb \equiv
[\rho(t,{\bf x}) - \rho_b(t)]/\rho_b(t)$ rather than in the trajectories. Since the density contrast can be expressed in terms of the trajectories of the particles, it should be possible to write down a differential equation for $\delta (t, \bld x)$ based on the equations for the trajectories $\bld x_i (t)$ derived above. It is, however, somewhat easier to write down an equation for $\delta_{\bld k} (t)$ which is the spatial Fourier transform of $\delta (t, \bld x)$. To do this, we begin with the fact that the density $\rho(\bld x,t)$ due to a set of point particles, each of mass $m$, is given by
\begin{equation}
\rho (\bld x,t) = {m\over a^3 (t)} \sum\limits_i \delta_D [ \bld x - \bld x _i(t)]
\end{equation}
where $\bld x_{i}(t)$ is the trajectory of the ith particle. To verify the $a^{-3}$ normalization, we can calculate the average of $\rho(\bld x,t)$ over a large volume $V$. We get
\begin{equation}
\rho_b(t) \equiv \int  {d^3 \bld x \over V} \rho (\bld x, t) = {m\over a^3(t)} \lb {N\over V}\rb = {M\over a^3 V} = {\rho_0\over a^3}
\end{equation}
where $N$ is the total number of particles inside the volume $V$ and $M = Nm$ is the mass contributed by them. Clearly $\rho_b \propto a^{-3}$, as it should.  The density contrast $\delta (\bld x,t)$ is related to $\rho(\bld x, t)$ by
\begin{equation}
1+\delta (\bld x,t) \equiv {\rho(\bld x, t) \over \rho_b} = {V \over N} \sum\limits_i \delta_D [\bld x - \bld x_i(t)] =  \int d {\bld q} \delta_D [\bld x - \bld x_{T} (t, \bld q)]  . 
\end{equation}
In arriving at the last equality we have taken the continuum limit by replacing: (i) $\bld x_i(t)$ by $\bld x_T(t,\bld q)$ where  $\bld q$ stands for a set of parameters (like the initial position, velocity etc.) of a particle; for
simplicity, we shall take  this to be initial position.  (ii) $(V/N)$ by $d^3{\bld q}$ since both represent volume per particle. Fourier transforming both sides  we get
\begin{equation}
\delta_{\bld k}(t) \equiv \int d^3\bld x   {\rm e}^{-i\bld k \cdot \bld x} \delta (\bld x,t) =   \int d^3 {\bld q} \  {\rm exp}[ - i {\bf k} . {\bf x}_{T} (t, \bld q)]  -(2 \pi)^3 \delta_D (\bld k)
\end{equation}
Differentiating this expression, 
and using the equation of motion (\ref{twnine}) for the trajectories give, after straightforward algebra, the equation (see  \cite{pines}, \cite{tpprob}, \cite{lssu}): 
\begin{equation}
\ddot \delta_{\bf k} + 2 {\dot a \over a} \dot \delta_{\bf k} = {1\over a^2}\int d^3 \bld q 
e^{-i{\bf k}. {\bf x }_T(t, \bld q) }\left\{ i{\bf k}\cdot \nabla \phi - a^2 ({\bf k \cdot \dot  x}_T)^2\right\}
 \label{basic}
\end{equation}
which can be further manipulated to give
\begin{equation}
\ddot \delta_{\bf k} + 2 {\dot a \over a} \dot \delta_{\bf k} = 4 \pi G \rho_b \delta_{\bf k} + A _{\bld k}- B_{\bld k} \label{exev}
\end{equation}
with
\begin{equation} 
A_{\bld k} =4\pi  G\rho_b \int{d^3{\bf k}' \over (2 \pi)^3}  \delta_{\bf k'} \delta_{{\bf k} - {\bf k'}} \left[{{\bf k}. {\bf k'} \over k^{'2}} \right] 
\end{equation}
\begin{equation}
B_{\bld k} = \int d^3 \bld q    \left({\bf k}.{\dot{\bf x}_T}  \right)^2 {\rm exp} \left[ -i{\bf k}. {\bf x }_T(t, \bld q) \right] .\label{exevii} 
\end{equation}
This equation is exact but involves $\dot{\bf x}_{T}(t, \bld q)$  on the right hand side and hence cannot be considered as closed. 
 
The structure of (\ref{exev}) and (\ref{exevii}) can be simplified if we use the perturbed gravitational potential (in Fourier space) $\phi_{\bf k}$ related to $\delta_{\bf k}$ by 
\begin{equation}
\delta_{\bf k} = - {k^2\phi_{\bld k} \over 4 \pi G \rho_b a^2} = - \lb {k^2 a \over 4 \pi G \rho_0}\rb \phi_{\bld k} = - \lb {2 \over 3H_0^2 }\rb k^2a \phi_{\bld k}
\end{equation}
and write the integrand for $A_{\bld k}$ in the symmetrised form as 
\begin{eqnarray}
\delta_{\bld k'} \delta_{\bld k - \bld k'} \left[ {\bld k . \bld k' \over k^{'2}} \right]& = &{1 \over 2} \delta_{\bld k'} \delta_{\bld k - \bld k'}\left[  {\bld k . \bld k' \over k^{'2}}  + {\bld k . (\bld k - \bld k') \over | \bld k - \bld k'|^2} \right] \nonumber \\
&=& { 1\over 2} \left( {\delta_{\bld k}'} \over k^{'2} \right) \left( {\delta_{\bld k - \bld k'} \over | \bld k - \bld k'|^2} \right) \left[ (\bld k - \bld k')^2 \bld k . \bld k' + k^{'2}\left( k^2 - \bld k . \bld k'\right)\right]\nonumber \\
&=& {1\over 2} \left({2a \over 3H_0^2}\right)^2 \phi_{\bld k'} \phi_{\bld k - \bld k'} \left[ k^2 (\bld k . \bld k' + k^{'2}) - 2(\bld k . \bld k')^2 \right] \nonumber \\
\end{eqnarray}
In terms of $\phi_{\bld k}$ (with ${\bf k}' = ({\bf k}/2) +{\bf p})$, equation (\ref{exev}) becomes, 
\begin{eqnarray}
\ddot \phi_{\bf k} + 4 {\dot a \over a} \dot\phi_{\bf k}   &= & - {1 \over 2a^2} \int {d^3{\bf p} \over (2 \pi )^3} \phi_{{1\over 2}{\bf k+p}}   
\phi_{{1\over 2}{\bf k-p}}\left[ \left( {k\over 2}\right)^2 + p^2 
-2  \lb {\bld k . \bld p \over k}\rb^2 \right] \nonumber \\
&+ &\lb{3H_0^2 \over 2}\rb  \int {d^3{\bf q} \over a} \lb{\bld k} . \dot {\bld x}\over k\rb ^2 e^{i{\bf k}.{\bf x}} \label{powtransf} 
\end{eqnarray}
where $\bld x = \bld x_T(t, \bld q)$. 
We shall now consider several applications of this equation.

  \subsection{\label{renormgrav}Application 1:  `Renormalizability' of gravity\index{Renormalizability of gravity}}
  
  Gravitational clustering in an expanding universe brings out an interesting feature 
  about gravity which can be described along the following lines. Let us consider a large
  number of particles which are  interacting via gravity in an expanding background and form
  bound gravitating systems. At some time $t$, let us assume that a fraction $f$ of the particles are
  in virialized, self-gravitating clusters which are reasonably immune to the effect of 
  expansion. Imagine that we replace each of these clusters by  single particles
  at their  centers of mass with masses equal to the total mass of the corresponding clusters. 
  (The total number of 
  particles have now been reduced but, if the original number was sufficiently large, we may
  assume that the resulting number of particles is again large enough to carry on further evolution
  with a valid statistical description.) We now evolve the resulting system to a time $t'$ and 
  compare the result with what would have been obtained if we had evolved the original system
  directly to $t'$. Obviously, the characteristics of the system at small scales (corresponding to the
  typical size $R$ of the clusters at time $t$) will be quite different.
  However, at large scales ($kR\ll 1$), the characteristics will be the same both the systems. In other words,
  the effect of a bunch of particles, in a virialized cluster, on the rest of the system
  is described, to the lowest order, by just the monopole moment of the cluster --
  which is taken into account by replacing the cluster by a single particle at the center of mass
  having appropriate mass. In this sense, gravitational interactions are ``renormalizable'' --
  where the term is used in the specific sense defined above.
  
  The result has been explicitly verified in simulations \cite{tp1} but one must emphasize that
  the whole idea of numerical simulations of such systems tacitly assumes the validity of this 
  result. If the detailed non linear behaviour at small scales, say within galaxies, influences
  very large scale behaviour of the universe (say, at super cluster scales), then it will be
  impossible to simulate large scale structure in the universe with finite resolution.
 
  One may wonder how this feature (renormalizability of gravity) is taken care
  of in equation (\ref{exev}). Inside a galaxy cluster, for example, the velocities
  ${\bf \dot x}_T$  can be quite high and one might think that this could influence
  the evolution of $\delta_{\bf k}$ at all scales. This does not happen and, to the lowest order,
  the contribution from virialized bound clusters cancel in $A_{\bf k} - B_{\bf k}$.  
  We shall now provide a proof of this result \cite {lssu}. 
  
  We begin by  writing the right hand side ${\cal R}$ of the 
  equation (\ref{basic}) concentrating on the particles in a given cluster.
  \begin{equation}
 {\cal R}  = \int d^3 {\bf q}\,  e^{-i{\bf k\cdot Q}} i k^a \left\{ 
  {\partial_a \phi\over a^2} + i \dot Q^a ({\bf k \cdot \dot Q})\right\}
  \label{altone}
  \end{equation} 
  where we have used the notation ${\bf Q} = {\bf x}_T$ for the trajectories of the particles
  and the subscripts $a, b, .... = 1, 2, 3 $ denote the components of the vector.
  For a set of particles which form a bound virialized cluster, we have from (\ref{twnine}) the
  equation of motion 
  \begin{equation}
  \ddot Q^i + 2 {\dot a\over a} \dot Q^i = - {1\over a^2} {\partial \phi\over \partial Q^i}
  \end{equation}
  We multiply this equation by
  $Q^j $, sum over the particles in the particular cluster and symmetrize on $i $ and $j$, to obtain
  the equation 
  \begin{equation}
  {d^2\over dt^2} \sum Q^i Q^j - 2 \sum \dot Q^i \dot Q^j + 2 {\dot a\over a} {d\over dt}
  \sum Q^iQ^j = -{1\over a^2} \sum \left( Q^j {\partial \phi \over \partial Q^i} + Q^i {\partial \phi\over 
  \partial Q^j}\right)
  \label{alttwo}
  \end{equation}
 We use the summation symbol, rather than integration over ${\bf q}$ merely to emphasize the fact that the sum is over particles of a given cluster. Let us now consider the first term in the right hand side of  (\ref{altone}) with the origin
  of the coordinate system shifted to the center of mass of the cluster. Expanding the exponential
  as $e^{-i{\bf k\cdot Q}} \approx (1-i{\bf k\cdot Q}) + {\cal O}(k^2 R^2) $ where $R$
  is the size of the cluster, we find that in the first term,  proportional to $\nabla \phi$, 
    the sum of the forces acting on
  all the particles in the cluster (due to self gravity) vanishes. The second term gives, on symmetrization,
  \begin{equation}
  \sum i a^{-2}({\bf k\cdot }\nabla \phi)e^{-i{\bf k\cdot Q}} \approx {k^ak^b\over 2a^2} \sum \left( Q^b 
  {\partial \phi\over \partial Q^a} + Q^a  {\partial \phi\over \partial Q^b}\right)
  \end{equation}
  Using (\ref{alttwo}) we find that 
  \begin{equation}
   \sum i a^{-2}({\bf k\cdot }\nabla \phi)e^{-i{\bf k\cdot Q}} = + \sum ({\bf k\cdot \dot Q})^2 + {1\over 2}
   \left( {d^2 \over dt^2} + 2 {\dot a\over a} {d\over dt}\right) \sum ({\bf k\cdot Q})^2
   \end{equation}
   The second term is of order ${\cal O}(k^2R^2)$ and can be ignored, giving
   \begin{equation}
   \sum i a^{-2}({\bf k\cdot }\nabla \phi)e^{-i{\bf k\cdot Q}} \approx + \sum ({\bf k\cdot \dot Q})^2
    +{\cal O}(k^2R^2)
   \end{equation}
   Consider next the second term in the right hand side of (\ref{altone}) with the same
   expansion for the exponential. We get
   \begin{eqnarray}
   \sum (ik^a) e^{-i{\bf k\cdot Q}} &\left[ i \dot Q^a k^b \dot Q^b\right] \cong - \sum k^ak^b \dot Q^a
   \dot Q^b ( 1- i {\bf k\cdot Q}) +{\cal O}(k^2R^2)\nonumber \\
   &= - \sum ({\bf k\cdot \dot Q})^2 + \sum ({\bf k\cdot \dot Q})^2 k^a Q^a +{\cal O}(k^2R^2)
   \end{eqnarray}
   The second term is effectively zero for any cluster of particles for which ${\bf Q}\to -{\bf Q}$
   is a symmetry. Hence the two terms on the right hand side of (\ref{altone}) cancel each other
   for all particles in the same virialized cluster; that is, the term $(A_{\bf k} - B_{\bf k})$ receives
   contribution only from particles which are not bound to any of the clusters to the order
   ${\cal O}(k^2R^2)$. If the typical size of the clusters formed at time $t$
   is $R$, then for wave-numbers with $k^2 R^2 \ll 1$, we can ignore the contribution from the clusters.
   Hence, in the limit of $k \to 0$ we can ignore $(A_{\bf k} -B_{\bf k})$
   term and treat equation (\ref{exev}) as linear in $\delta_{\bf k}$; 
    large spatial scales in the universe can be described by linear perturbation theory
   even when small spatial scales are highly non linear.
   
   There is, however, an important caveat to this claim. In the right hand side of (\ref{exev}) one 
   is comparing the first term (which is linear in $\delta_{\bf k}$) with  the contribution
   $(A_{\bf k} -B_{\bf k})$. If, at the relevant wavenumber, the first term  $4\pi G \rho_b \delta_{\bf k}$
   is negligibly small, then
   the {\it only} contribution will come from $(A_{\bf k} -B_{\bf k})$ and, of course, we cannot
   ignore it in this case. The above discussion shows that this contribution will scale as 
   $k^2 R^2$ and will lead to a development of $\delta_{\bf k} \propto k^2$ if originally (in linear 
   theory) $\delta_k \propto k^n$ with $n>2$ as $k\to 0$.  
   
   We shall next describe 
    the linear evolution; the development of $\delta_{\bf k} \propto k^2$ tail
   at large spatial scales will be taken up in section \ref{nltail}.

\subsection{\label{evollarge}Application 2: Evolution at  large scales.}
 
If the density contrasts are small and linear perturbation theory is to be valid, we should be able to  ignore the terms $A_{\bld k}$ and $B_{\bld k}$ in (\ref{exev}). Thus the linear perturbation theory in Newtonian limit is governed by the equation
\begin{equation}
\ddot \delta_{\bf k} + 2 {{\dot a} \over a} \dot \delta_{\bf k} = 4 \pi G \rho_b \delta_{\bf k} \label{linpertb}
\end{equation} 
From the structure of equation (\ref{exev}) it is clear that we will obtain the linear equation if $A_{\bld k} \ll 4 \pi G\rho_b\delta_{\bld k}$ and $\bld B_{\bld k} \ll 4 \pi G \rho_b \delta_{\bld k}$. A {\it necessary} condition for this $\delta_{\bld k} \ll 1$ but this is {\it not} a sufficient condition --- a fact often ignored or incorrectly treated in literature. As we saw in the last section, if $\delta_{\bld k} \rightarrow 0$ for certain range of $\bld k$ at $t = t_0$ (but is nonzero elsewhere) then $(A_{\bld k}-B_{\bf k}) \gg 4 \pi G \rho_b \delta_{\bld k}$ and the growth of perturbations around $\bld k$ will be entirely determined by nonlinear effects.

 For the present, we shall assume that $A_{\bld k}-B_{\bld k}$ is ignorable and study the resulting system.
In that case, equation (\ref{linpertb}) has the growing solution $\delta_{\bf k}(t) = [a(t)/a(t_i)] \delta_{\bf k}(t_i)$ in the matter dominated universe with $a(t)\propto t^{2/3},\rho_b\propto a^{-3}$. This shows that when linear perturbation theory is applicable the density perturbations grow as 
$a(t)$. The power spectrum $P({\bf k},t) = |\delta_{\bf k}(t)|^2$ and the correlation function
$\xi({\bf x},t)$ [which is the Fourier transform of the power spectrum] both grow as $a^2(t)$ 
while $\phi_{\bf k} \propto k^{-2} (\delta_{\bf k} /a )$ remains constant in time.
This analysis allows us to fix the evolution of clustering at sufficiently large scales
(that is, for sufficiently small $k$) uniquely. The clustering at these scales 
which is well described by linear theory, and the power spectrum  grows as $a^2$.

\subsection{\label{firstnl}Application 3: Formation of first  non linear structures\index{non linear structures}}

A useful insight into the nature of linear perturbation theory (as well as nonlinear clustering) can be
obtained by examining the nature of particle trajectories which lead
to the growth of the density contrast $\delta_L (a) \propto a$ obtained above.
To determine the particle trajectories corresponding to the 
linear limit, let us start by writing the trajectories in the form
\begin{equation} 
{\bf x}_T (a,{\bf q}) = {\bf q} + {\bf L} (a,{\bf q})
\end{equation}
where ${\bf q}$ is the Lagrangian coordinate (indicating the 
original position of the particle) and ${\bf L}(a,{\bf q})$ is 
the displacement. The corresponding Fourier transform of the density contrast is given by the general expression
\begin{equation}
 \delta_{\bf k} (a)= \int d^3{\bf q}\, e^{-i{\bf k\cdot q}-i{\bf k\cdot L}(a,{\bf q})} - (2 \pi)^3 \delta_{\rm Dirac} [{\bf k}]
\end{equation}
In the linear regime, we expect the particles to have moved very little
and hence we can expand the integrand in the above equation in a Taylor
series in $({\bf k\cdot L})$. This gives, to the lowest order, 
\begin{equation} 
\delta_{\bf k} (a)\cong -\int d^3{\bf q}\, e^{-i{\bf k\cdot q}} (i{\bf k\cdot L}(a,{\bf q})) = -\int d^3{\bf q}\, e^{-i{\bf k\cdot q}}\lb \nabla_{\bf q} \cdot {\bf L}\rb
\end{equation}
showing that $\delta_{\bf k}(a)$ is Fourier transform of $-\nabla_{\bld q} . \bld L (a, \bld q)$. This allows  us to identify $\nabla\cdot {\bf L}(a,{\bf q})$ with
the original density contrast in real space $- \delta_{\bf q} (a)$. Using
the Poisson equation  we can write $\delta_{\bf q}(a)$ as a divergence; that is 
\begin{equation} 
\nabla \cdot {\bf L}(a,{\bf q}) = - \delta_{\bf q}(a) = - \fra{2}{3} H_0^{-2} a \nabla \cdot (\nabla \phi)
\end{equation}
which, in turn, shows that    a consistent set of displacements that will
lead to $\delta (a) \propto a$ is given by
\begin{equation}
{\bf L}(a,{\bf q}) = -  (\nabla \psi)a \equiv a {\bf u}({\bf q}) ; 
   \qquad \psi\equiv (2/3) H_0^{-2}\phi \label{ninety}
\end{equation} 
The trajectories in this limit
are, therefore, linear in $a$: 
\begin{equation}
 \bld x_{T} (a,{\bf q}) = {\bf q} + a {\bf u}({\bf q})\label{trajec}
\end{equation} 

A useful approximation to describe the quasi linear stages of clustering is obtained by using the trajectory in (\ref{trajec})  as an ansatz valid {\it even at quasi linear epochs}. In this approximation, called  Zeldovich approximation\index{Zeldovich approximation}, the proper Eulerian position $\bld r $ of a particle is related to its Lagrangian position $\bld q $ by 
\begin{equation}
{\bf r}(t) \equiv a(t) {\bf x}(t) = a(t) [{\bf q} +  
a(t) {\bf u}({\bf q}) ] \label{lagq}  
\end{equation}
where ${\bf x}(t)$ is the comoving Eulerian coordinate.
 If the initial, unperturbed, 
density is $\overline \rho$ (which is independent of ${\bf q})$,
then the conservation of mass implies that the perturbed density will be
\begin{equation}
\rho ({\bf r},t) d^3{\bf r} = \bar \rho d^3{
\bf q}.\label{qmcons}
\end{equation}
Therefore
\begin{equation}
\rho({\bf r},t) = \bar \rho  \left[{\rm det} \lb{ \partial q_i \over \partial r_j}\rb\right]^{-1} = 
{\bar \rho/a^3 \over {\rm det}
 (\partial x_j/\partial q_i)} = {\rho_b(t) 
\over {\rm det}
( \delta_{ij} + a(t) (\partial u_j/\partial q_i))}\label{qjacob}
\end{equation}
where we have set $\rho_b(t) = [\bar \rho / a^3(t)]$.
Since ${\bf u}({\bf q})$ is a gradient of a scalar function, 
the Jacobian in the denominator of (\ref{qjacob}) is the determinant of a real symmetric
matrix. This matrix 
can be diagonalised at every point ${\bf q}$, to yield a set of
eigenvalues and principal axes as a function of ${\bf q}$. 
If the eigenvalues of $(\partial u_j/
\partial q_i) $ are $[-\lambda_1({\bf q})$, $-\lambda_2({\bf q})$, 
$-\lambda_3({\bf q})]$ then the perturbed density is given by
\begin{equation}
\rho({\bf r},t) = {\rho_b(t) \over (1 - a(t)\lambda_1({\bf q}))
(1 - a(t) \lambda_2({\bf q}))
(1 - a(t)\lambda_3({\bf q}))} \label{qeig}
\end{equation}
where ${\bf q}$ can be expressed as a function of ${\bf r}$ by solving (\ref{lagq}).
This expression describes the effect of
deformation of an infinitesimal, cubical,
volume (with the faces of the cube
determined by the eigenvectors corresponding to $\lambda_n$)
and the consequent change in the density. 
A positive $\lambda$
denotes collapse and negative $\lambda$
signals expansion.

In a over dense region, the density will become 
infinite if one of the terms in brackets in the denominator of (\ref{qeig}) 
becomes zero. In the generic case,
these eigenvalues will be different
from each other;
so that we can take, say,  $\lambda_1\geq \lambda_2\geq \lambda_3$. 
At any particular value  of ${\bf q}$ the density  will  diverge for the first time when 
$(1 - a(t)\lambda_1) = 0$;
at this instant 
 the material contained in a cube in the 
${\bf q}$ space gets compressed to a sheet in the ${\bf r}$ space, 
along the principal axis corresponding to $\lambda_1$.
Thus sheet like structures, or `pancakes'\index{pancakes}, will
be the first nonlinear structures to form when gravitational instability
amplifies density perturbations.

  \subsection{\label{nltail}Application 4: A  non linear tail at small wavenumber}
  
  There is an interesting and curious result which is characteristic of gravitational
  clustering that can be obtained directly from our equation (\ref{powtransf}). Consider an initial
  power spectrum which has very little power at large scales; more precisely, we shall
  assume that $P(k)$ dies faster than $k^4$ for small $k$. If these large scales are described
  by linear theory --- as one would have normally expected, then the power at these scales 
  can only grow as $a^2$ and it will always be sub dominant to $k^4$. It turns out that this 
  conclusion is incorrect.   
As the system evolves, small scale nonlinearities will
develop in the system and --- if the large scales have too little
power intrinsically (i.e. if $n$ is large) ---  then
the long wavelength power will soon be dominated by the
``tail'' of the short wavelength power arising from the
nonlinear clustering. This occurs because, in equation (\ref{exev}), the nonlinear term 
 ($A_{\bld k}$ - $B_{\bld k} )= {\cal O}(k^2R^2)$ can dominate over $4 \pi G \rho_b\delta_{\bld k}$ at long wavelengths  (as ${\bld k} \rightarrow 0$) and   lead to the development of a $k^4$ power spectrum at
  large scales. This is a purely non linear effect which we shall now describe.
  
  To do this, we shall use the Zeldovich approximation to obtain \cite{tpiran} a closed equation 
  for $\phi_{\bf k}$.
The trajectories in Zeldovich approximation, given by  (\ref{trajec}) can be used in (\ref{powtransf}) to provide a {\it closed} integral equation for $\phi_{\bld k}$. In this case,
\begin{equation}
\bld x_T(\bld q, a) = \bld q + a \nabla \psi ; \quad \dot \bld x_{\rm T} = \lb {2a \over 3t}\rb \nabla \psi; \quad \psi = {2 \over 3H_0^2 } \varphi
\end{equation}
and, to the same order of accuracy, $B_{\bld k}$ in (\ref{exevii}) becomes:
\begin{equation}
\int d^3 \bld q \lb \bld k \cdot \dot\bld x_{\rm T}\rb^2e^{-i \bld k \cdot(\bld q + \bld L)} \cong \int d^3 \bld q ( \bld k \cdot \dot \bld x_{\rm T})^2 e^{-i \bld k \cdot \bld q}
\end{equation}
Substituting these expressions in (\ref{powtransf}) we find that the gravitational potential is described by the closed integral equation:
\begin{eqnarray}
\ddot \phi_{\bld k} + 4 {\dot a \over a} \dot \phi_{\bld k} &=& -{1 \over 3a^2} \int {d^3 \bld p \over (2 \pi)^3} \phi_{{1 \over 2} \bld k + \bld p} \phi_{{1 \over 2} \bld k - \bld p} {\cal G} (\bld k, \bld p)\nonumber \\
{\cal G} (\bld k, \bld p) &= &{7 \over 8} k^2 + {3 \over 2} p^2 - 5 \lb {\bld k \cdot \bld p\over k}\rb^2 \label{calgxx} \nonumber \\
\end{eqnarray}
This equation provides a powerful method for analyzing non linear clustering since estimating $(A_{\bld k}-B_{\bld k})$ by Zeldovich approximation has a very large domain of applicability.

A  formal way of obtaining the $k^4$ tail is to solve equation (\ref{calgxx}) for long wavelengths  \cite{tpiran}; i.e. near $\bld k = 0$.   Writing $\phi_{\bld k} = \phi_{\bld k}^{(1)} + \phi_{\bld k}^{(2)} + ....$ where $\phi_{\bld k}^{(1)} = \phi_{\bld k}^{(L)}$ is the time {\it independent} gravitational potential in the linear theory and $\phi_{\bld k}^{(2)}$ is the next order correction, we get from (\ref{calgxx}), the equation  
\begin{equation}
\ddot\phi_{\bld k}^{(2)}+ 4 {\dot a \over a} \dot\phi_{\bld k}^{(2)} \cong - { 1 \over 3a^2} \int {d^3 \bld p \over (2 \pi)^3} \phi^L_{{1 \over 2} \bld k + \bld p} 
\phi^L_{{1 \over 2} \bld k - \bld p} {\cal G}(\bld k, \bld p)
\end{equation}
The solution to this equation is the sum of a solution to the homogeneous part [which decays as 
$\dot\phi\propto a^{-4}\propto t^{-8/3}$ giving $\phi\propto t^{-5/3}$] and a particular solution which grows as $a$. Ignoring the decaying mode at late times and taking
$\phi_{\bld k}^{(2)} = aC_{\bld k}$ one can determine $C_{\bld k}$ from the above equation. Plugging it back, we find the lowest order correction to be,
\begin{equation}
\phi_{\bld k}^{(2)} \cong - \lb {2a \over 21H^2_0}\rb \int {d^3 \bld p \over (2 \pi)^3}\phi^L_{{1 \over 2} \bld k + \bld p} 
\phi^L_{{1 \over 2} \bld k - \bld p} {\cal G}(\bld k, \bld p)
\label{approsol}
\end{equation}
Near $\bld k \simeq 0$, we have
\begin{eqnarray}
\phi_{\bld k \simeq 0}^{(2)} &\cong& - {2a \over 21H^2_0} \int {d^3 \bld p \over (2 \pi)^3}|\phi^L_{\bld p}|^2 \left[ {7 \over 8}k^2 + {3 \over 2}p^2 - {5(\bld k \cdot \bld p)^2 \over k^2} \right] \nonumber \\
&=&   {a \over 126 \pi^2H_0^2} \int\limits^{\infty}_0 dp  p^4 |\phi^{(L)}_{\bld p}|^2\nonumber \\
\end{eqnarray}
which is independent of $\bld k$ to the lowest order. Correspondingly the power spectrum for
 density $P_{\delta}(k)\propto a^2 k^4P_{\varphi} (k) \propto a^4 k^4$ in this order. 

The generation of long wavelength $k^4$ tail is easily seen in simulations if one starts with a power spectrum that is sharply peaked in $|\bld k|$. Figure \ref{figptsimu}  shows the results of such a simulation \cite{jsbtp1}  in which the y-axis is $[\Delta(k)/a(t)]$
where $\Delta^2(k) \equiv k^3P/2\pi^2$ is the power per logarithmic band in $k$. 
 In linear theory $\Delta \propto a$ and this quantity should not change. The curves labelled by $a=0.12$ to $a=20.0$ show the effects of nonlinear evolution, especially the development of $k^4$ tail.

%%%%%%%%%% Figure  %%%%%%%%%%%%%%%%%%%%%%
\begin{figure}[ht]
\begin{center}
\includegraphics[width=.8\textwidth]{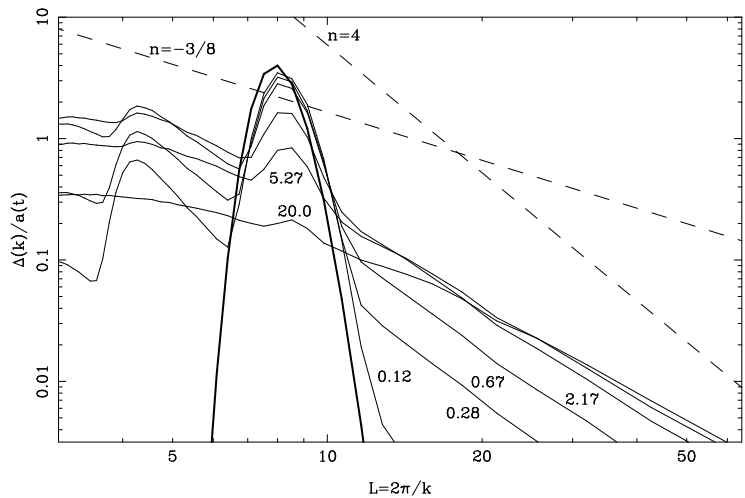}
\end{center}
\caption{The transfer of power to long wavelengths forming a $k^4$ tail is illustrated using 
simulation results. Power is injected in the form of a narrow peak at $L=8$. 
Note that the $y-$axis is $(\Delta/a)$ so that
there will be no change of shape under linear evolution
with $\Delta\propto a$. As time goes on a $k^4$ tail is generated which
 itself evolves according to the nonlinear scaling relation discussed later on}
\label{figptsimu}
\end{figure}
%%%%%%%%%% End   Figure  %%%%%%%%%%%%%%%%%%%%%%

   %%%%%%%%%%%%%%%%%%%%%%%%%%%%%%%%%%%%% 
     \begin{figure}[ht]
   \begin{center}
   \includegraphics[width=0.85\textwidth]{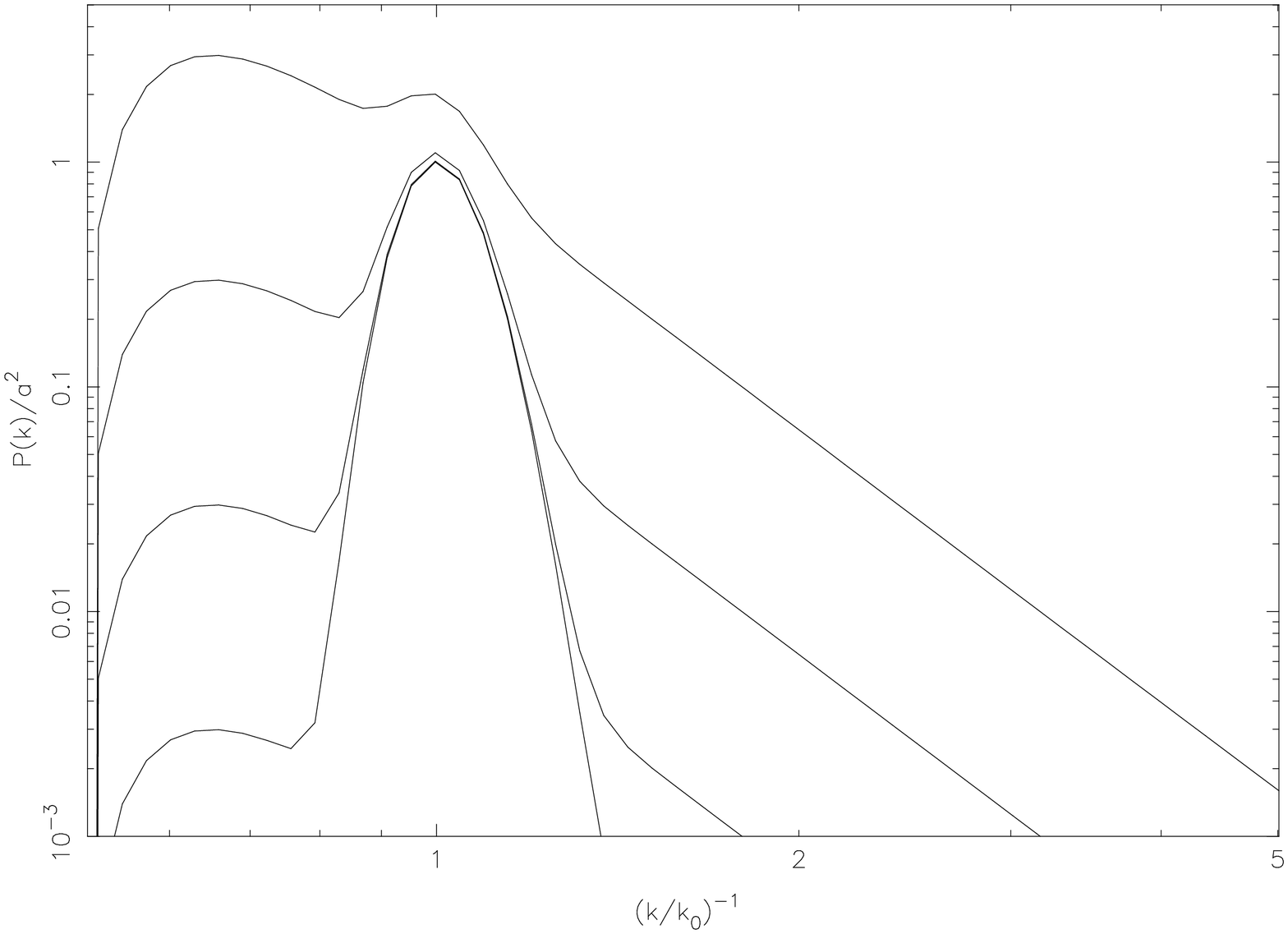}
   \end{center}
   \caption{Analytic model for transfer of power in gravitational clustering. The initial power
    was injected at the wave number $k_0$ with a Gaussian window of width $\Delta k/k_0= 0.1$.
    First order calculation shows that the power is transfered to larger spatial scales with a 
    $k^4$ tail and to the shorter spatial scales, all the way down to  $(1/2)k_0^{-1}$. The plot gives the
    total power spectrum divided by $a^2$ (with $y-$axis normalized arbitrarily) at different times with $a^2$
    changing by factor 10  between any two curves}
    \label{figurepttheory}
    \end{figure}
    
    %%%%%%%%%%%%%%%%%%%%%%%%%%%%%%%%%%

\subsection{\label{gensmall}Application 5: Generation of small scale power}

Figure \ref{figptsimu}   also shows that, as the clustering proceeds, power is generated at  spatial
scales smaller than the scale $k_0^{-1}$ at which the power is injected. This feature can 
also be easily understood \cite{tp1} from our equation (\ref{approsol}). Let the initial gravitational potential and
the density contrast 
(in the linear theory)  be sharply peaked at the wave number $k_0$, say, with:
\begin{equation}
\phi^L_{\bf k} = \mu {H_0^2\over k_0^4} \delta_D [|{\bf k}| - k_0]; \quad k_0^3 \delta^L_{\bf k} =-{2\over 3}
(\mu a) k_0 \delta_D [|{\bf k}| - k_0] 
\label{initcond}
\end{equation}
where $\mu$ is dimensionless constant indicating the strength of the potential and the other factors ensure the correct dimensions. Equation (\ref{approsol}) shows that, the right hand side is nonzero only when the
{\it magnitudes} of both the vectors $[(1/ 2){\bf k}+{\bf p}]$ and $[(1/ 2){\bf k}-{\bf p}]$ are $k_0$. This
requires ${\bf k}\cdot{\bf p}=0, (k/2)^2+p^2=k_0^2$. 
(This constraint  has a simple geometric interpretation: Given any
${\bf k}$, with $k\le 2k_0$ one constructs a vector ${\bf k}/2$ inside a sphere of radius $k_0$ and
a vector ${\bf p}$ perpendicular to ${\bf k}/2$  reaching up to the shell at radius $k_0$
where the initial power resides. Obviously, this construction is possible only for
   $k< 2k_0$.) Performing the integration in (\ref{approsol})  we find that
\begin{equation}
\phi^{(2)}_{\bf k} = {\mu^2\over 56\pi^2} {H_0^2\over k_0^5} a \left( 1 - {k^2 \over 4 k_0^2}\right)
\left( 1 + {k^2\over 3 k_0^2}\right)
\end{equation}
[We have again ignored the decaying mode which arises as a solution to the homogeneous part.]
The corresponding power spectrum for the density field
 $P(k)= |\delta_k|^2 \propto a^2 k^4 |\phi_k|^2 $ will evolve as
\begin{equation}
P^{(2)} (k) \propto (\mu a)^4 q^4 \left( 1 - {1\over 4} q^2\right)^2 \left( 1 + {1\over 3} q^2\right)^2; \quad 
q= {k\over k_0}
\end{equation}
    The power at large spatial scales ($k\to 0$) varies as $k^4$ as discussed before.
    The power has also been generated at smaller scales in the range $k_0<k<2k_0$ with 
    $P^{(2)}(k)$  being a maximum at $k_m\approx 1.54 k_0$ corresponding to the length scale
    $k_m^{-1}\approx 0.65 k_0^{-1}$. 
    Figure \ref{figurepttheory}  shows
    the power spectrum for density field (divided by $a^2$ to eliminate linear growth)
     computed analytically  for a narrow Gaussian
    initial power spectrum centered at $k_0 =1$. The curves are for $(\mu a /56\pi^2)^2 =
    10^{-3}, 10^{-2}, 10^{-1}$ and 1. The similarity between figures \ref{figptsimu}  and
    \ref{figurepttheory}  
    is striking and allows us to understand the simulation results. The key difference
    is that, in the simulations, newly generated power will further produce power at
    $4k_0, 8k_0, ...$ and each of these will give rise to a $k^4$ tail to the right.
    The resultant power will, of course, be more complicated than predicted by our analytic 
    model.   
    The generation of power near this maximum at $k_m^{-1} = 0.65 k_0^{-1}$
     is clearly visible as a second peak
    in figure  \ref{figurepttheory}  and around 
    $2\pi/k_0\approx  4$ in figure \ref{figptsimu} . 
    
    If we had taken the initial power spectrum to be
    Dirac delta function in the wave vector ${\bf k}$ (rather than on the {\it magnitude} of the wave vector,
    as we have done) the right hand side of (\ref{approsol}) will contribute
    only when  $({1\over 2}{\bf k}\pm{\bf p}) = {\bf k}_0$. This requires ${\bf p} =0$ and ${\bf k} = 2{\bf k}_0$
    showing that the power is generated exactly at the second harmonic of the wave number.
    Spreading the initial power on a shell of radius $k_0$, spreads the power over different vectors
    leading to the result obtained above.

    Equation (\ref{initcond}) shows that $k_0^3\delta_{\bf k}$ will reach  non linearity for
    $\mu a \approx (3/2)$. The situation is different as regards the gravitational potential
    due to the large numerical factor $56 \pi^2$; the gravitational potential fluctuations are 
    comparable to the original fluctuations only when $\mu a \approx 56 \pi^2$.
    We shall say more about power transfer in gravitational clustering in section \ref{cipt}.

\subsection{\label{sphapprox}Application 6:  Spherical approximation\index{Spherical approximation}}
 
In the nonlinear regime --- when $\delta\ga 1$ --- it is not possible to solve equation   (\ref{exev})  exactly. Some progress, however, can be made if we assume that the trajectories are homogeneous; i.e. $ \bld x (t, \bld q) = f (t)\bld q $ where $f(t)$ is to be determined. In this case, the density contrast is
\begin{equation}
\delta_{\bld k} (t)  = \int d^3 \bld q e^{-if(t)\bld k . \bld q} - (2 \pi)^3 \delta_D(\bld k)
 =(2\pi)^3 \delta_D (\bld k) [f^{-3} - 1] \equiv (2 \pi)^3 \delta_D (\bld k)\delta (t) \label{spheapprox}
\end{equation}
where we have defined $\delta(t) \equiv \left[ f^{-3}(t)-1 \right]$ as the amplitude of the density contrast for the $\bld k = 0$ mode. It is now straightforward to compute $A$ and $B$ in (\ref{exev}).  We have
\begin{equation}
A = 4 \pi G\rho_b \delta^2(t) [(2  \pi)^3 \delta_D(\bld k)] 
\end{equation}
and 
\begin{eqnarray} 
B&=&\int d^3 \bld q (k^aq_a)^2 \dot f^2 e^{-if(k_aq^a)} = -\dot f^2 {\partial^2 \over \partial f^2} [(2 \pi)^3 \delta _D(f \bld k) ] \nonumber \\
 &=& -{4 \over 3} {\dot \delta^2 \over (1 + \delta)} [(2 \pi)^3 \delta_D (\bld k)]
\end{eqnarray}
so that the equation (\ref{exev})  becomes 
\begin{equation}
\ddot\delta + 2 {\dot a \over a} \dot\delta = 4 \pi G \rho_b (1 + \delta) \delta + {4 \over 3} {\dot\delta^2 \over (1 + \delta)} \label{x}
\end{equation}
To understand what this equation means, let us consider, at some initial epoch $t_i$, a spherical region of the universe which has a slight constant over-density compared to the background. As the universe expands, the over-dense region will expand more slowly compared to the background, will reach a maximum radius, contract and virialize to form a bound nonlinear system. Such a model is  called ``spherical top-hat''.
It is possible to show that equation (\ref{x}) is the same as the equation governing density evolution in 
a spherical model \cite{tpiran}.  . The detailed analysis of the spherical model \cite{modsph} shows that
the virialized systems formed at any given time have a mean density which is typically 200 times the
background density of the universe at that time. Hence, it is often convenient to divide the growth of structures
into three regimes: linear regime with $\delta \ll 1$; quasi-linear regime with $1\la \delta\la 200$; and nonlinear
regime with $\delta\ga 200$.

\section{\label{nlscales} Nonlinear scaling relations\index{Nonlinear scaling relations}}

Given an initial density contrast, one can trivially obtain the two point
correlation function  at any later epoch in the {\it linear} theory. If there is a procedure for relating the nonlinear correlation function and linear correlation function (even approximately) then one can make considerable progress in understanding nonlinear clustering. It is actually possible to do this 
\cite{hklm}, \cite{rntp}, \cite{tpmodel}, \cite{tpnk} 
and relate the exact mean correlation function 
\begin{equation}
\bar\xi (t,x) = {3\over x^3} \int_0^x \xi(t,y) y^2 dy
\end{equation}
to the one computed in the linear theory,
 along the following lines:

The mean number of neighbours within a distance $x$ of any given particle is 
\begin{equation}
N(x,t)=(na^3)\int^{x}_{o}4\pi y^2dy[1+\xi(y,t)]\label{qmean}
\end{equation}
when $n$ is the comoving number density. Hence the conservation law for pairs implies 
\begin{equation}
{\partial\xi\over\partial t}+{1\over ax^2}{\partial\over \partial x}[x^2(1+\xi)v]=0\label{qcons}
\end{equation}
where $v(t,x)$ denotes the mean relative velocity of pairs at separation $x$ and
epoch $t$. Using
\begin{equation}
(1+\xi)={1\over 3x^2}{\partial\over \partial x}[x^3 (1+\bar{\xi})] 
\end{equation}
in (\ref{qcons}), we get
\begin{equation}
{1\over 3x^2}{\partial\over \partial x}[x^3{\partial\over\partial
t}( 1+\bar{\xi})] = - {1\over ax^2}{\partial\over \partial
x}\left[ {v\over 3} {\partial\over \partial x}[x^2(1+\bar{\xi})]\right].
\end{equation}
Integrating, we find:
\begin{equation}
x^3 {\partial\over \partial
t}(1+\bar{\xi})=-{v\over a}{\partial\over \partial x}[x^3(1+\bar{\xi})]. 
\end{equation}
[The integration would allow the addition of an arbitrary function of $t$ on
the right hand side. We have set this function to zero so as to reproduce
the correct limiting behaviour].
It is now convenient to change the variables from $t$ to $a$, thereby
getting an equation for $\bar{\xi}$:
\begin{equation}
a{\partial\over \partial
a}[1+\bar{\xi}(a,x)]=\left({v\over -\dot{a}x}\right)
{1\over x^2}{\partial\over \partial x}[x^3(1+\bar{\xi}(a,x))]
\label{qlim}
\end{equation}
or, defining $ h(a,x) = - (v/\dot{a}x)$
\begin{equation}
\left({\partial\over \partial \ln a}-h{\partial\over \partial \ln x}\right)\,\,\, (1+\bar{\xi})=3h \left( 1+\bar\xi\right).
\label{qhsi}
\end{equation}
This equation shows that the behaviour of $\bar{\xi}(a,x)$ is essentially
decided by $h$, the dimensionless ratio between the mean relative velocity $v$ and the Hubble velocity
$\dot{a}x=(\dot{a}/a)r$, both evaluated at scale $x$.

In the extreme nonlinear limit $\lb \bar \xi \gg 1 \rb$, we may expect bound structures  not to expand with Hubble flow. To maintain a stable structure, the relative pair velocity $v  \lb a, x \rb$ of particles separated by $x$ should balance the Hubble velocity $Hr = \dot a x;$ hence, $v  = - \dot a x $ or $h \lb a, x \rb \cong  1 $.
The behaviour of $h \lb a, x \rb$ for $\bar \xi \ll 1 $ is more complicated and can be 
shown \cite{tpiran},\cite{tpprob} that $h = (2/3) \bar\xi$ in the limit of $\bar\xi \ll 1$. 
Thus $h(a,x)$ depends on $(a, x)$ only through   $\bar\xi(a, x)$ in the linear limit, while $h \cong -1$ is the nonlinear limit. 
This suggests the ansatz that $h$ depends on $a$ and $x$ only through  $\bar\xi (a,x)$; that is, we assume that $h(a,x) = h [\bar\xi (a,x)]$.
 It is then  possible to find a solution to 
(\ref{qhsi}) 
which reduces to the form $\bar{\xi}\propto a^2$ for $\bar{\xi} \ll 1$ using the method of characteristics 
\cite{rntp}.
The final result can be given as:
\begin{equation}
\bar{\xi}_L(a,l)=\exp \left(
{2\over 3}\int^{\bar{\xi}(a,x)}{d\mu\over h(\mu)(1+\mu)}\right);\quad l=x[1+\bar{\xi}(a,x)]^{1/3}. \label{xibarint}
\end{equation}
Given the function $h(\bar{\xi})$, this relates $\bar{\xi}_{L}$ and
$\bar{\xi}$ or --- equivalently --- gives the mapping $\bar\xi(a,x)=U[\bar\xi_L(a,l)]$ between the nonlinear and linear correlation functions evaluated at different scales $x$ and $l$. The lower limit of the integral is chosen to give $\ln \bar \xi$ for small values of $\bar \xi$ on the linear regime.  [The $(2/3)$ factor in the exponent
becomes $(2/D)$ in D-dimensions and  $l=x[1+\bar{\xi}(a,x)]^{1/D}$.]

The following points need to be stressed regarding this result: (i) Among all statistical indicators, it is {\it only} $\bar\xi$ which obeys a nonlinear scaling relation (NSR) of the form $\bar\xi_{\rm NL} (a, x) = U\left[ \bar\xi_L(a,l) \right]$. Attempts to write similar relations for $\xi$ or $P(k)$ have no fundamental justification. (ii) The non locality of the relation represents the transfer of power in gravitational clustering and cannot be ignored --- or approximated by a local relation between $\bar\xi_{NL}(a,x)$ and $\bar\xi_L(a,x)$.

Given the form of $h(\bar\xi)$, equation (\ref{xibarint}) determines the relation $\bar\xi= U[\bar\xi_L]$. 
It is, however, easier to determine the form of $U$, directly from theory and this was
done in \cite{tpmodel}. Here, I shall provide a more intuitive but physically motivated derivation
along the following lines: 

In the linear regime $\lb \bar\xi \ll 1, \bar\xi_L \ll 1)\rb$ we clearly have $U(\bar\xi_L) \simeq \bar\xi_L$. 
We can divide the non linear phases of evolution conveniently into two parts, which I will call
quasi-linear ($1\la \bar\xi \la 200$) and non linear ($200 \la \bar\xi $). 
In the quasi-linear
phase, regions of high density contrast will undergo collapse and in the non linear phase
more and more virialized systems will get formed. 
We recall that, in the study of finite gravitating systems made of point particles and
interacting via Newtonian gravity, isothermal spheres play an important
role and are  the local maxima of entropy. Hence dynamical
evolution drives the system towards an $(1/x^2)$ density profile. Since one expects
similar considerations to hold at small scales, during the late stages of evolution of the universe, we may hope that isothermal spheres with
$(1/x^2)$ profile may still play a role in the late stages of evolution of 
clustering in an expanding background. However, while converting the density profile to correlation function,
 we need to distinguish between two cases. 
In the quasi-linear regime, dominated by the collapse of high density peaks,
the density profile around any peak will scale as the correlation function and
we will have $\bar\xi\propto (1/x^2)$. On the other hand, in the nonlinear
end, we will be probing the structure inside a single halo and $\xi({\bf x}) $ 
will vary as $<\rho({\bf x + y}) \rho({\bf y})>$. If $\rho \propto |x|^{-\epsilon}$, then $\xi \propto |x|^{-\gamma}$
with  $\gamma=2\epsilon -3$. This
gives $\bar\xi\propto (1/x)$ for $\epsilon=2$. Thus, if isothermal spheres
are the generic contributors, then we expect the correlation function to
vary as $(1/x)$ and nonlinear scales, steepening to $(1/x^2)$ at intermediate
scales. 
Further, since isothermal spheres are local maxima of entropy, a configuration like this should remain undistorted for a long duration. This
argument suggests that a $\bar\xi$ which goes as $(1/x)$ at small scales
and $(1/x^2)$ at intermediate scales is likely to  grow approximately as $a^2$ at all scales. 
The form of $U[\xi_L]$ must be consistent with this feature.

This criterion allows us to determine the form of $U$ in the quasi-linear and non linear phases.
In the quasi-linear regime, we want a linear correlation function of the form $\bar\xi_L
(a, l) = a^2/l^2$ to be mapped to $\bar\xi_{\rm NL} (a,x) \propto a^2/x^2$. We thus demand
  $U[a^2 l^{-2}] \propto a^2 x^{-2}$ with $l \approx x\xi_{\rm NL}^{1/3}$. It is trivial 
  to see that this requires $U(z) \propto z^3$. Similarly, in the non linear end,
  we expect
  a linear correlation function of the form $\bar\xi_L
  (a, l) = a^2/l$ to be mapped to $\bar\xi_{\rm NL} (a,x) \propto a^2/x$ requiring
  $U[a^2 l^{-1}] \propto a^2 x^{-1}$ with $l \approx x\xi_{\rm NL}^{1/3}$. This gives 
   $U(z) \propto z^{3/2}$.
   
  Combining  all the results  we find that the nonlinear mean
correlation function can be expressed in terms of the linear mean 
correlation function by the relation:
\begin{equation}
\bar \xi (a,x)=\cases{\bar \xi_L (a,l)&(for\ $\bar \xi_L<1, \, \bar
\xi<1$)\cr 
{\bar \xi_L(a,l)}^3 &(for\ $1<\bar \xi_L<5.85, \, 1<\bar \xi<200$)\cr
14.14 {\bar \xi_L(a,l)}^{3/2} &(for\ $5.85<\bar\xi_L, \, 200<\bar
\xi$)\cr}\label{hamilton} 
\end{equation}
The numerical coefficients have been determined by continuity
arguments. We have assumed the linear result to be valid up to
$\bar\xi=1$ and the 
virialisation to occur at $\bar\xi\approx 200$
which is result arising from the spherical model.  
The true test of such a model, of course, is N-body simulations and
remarkably enough, simulations are very well represented by relations
of the above form.  [The fact that numerical simulations show a correlation between
$\bar\xi(a,x)$ and $\bar\xi_L(a,l)$ was originally pointed out 
in \cite{hklm}.
  These authors, however, gave a multi 
 parameter
fit to the data which has the virtue of representing the numerical work accurately.
Equation (\ref{hamilton}), on the other hand, portrays the clear 
 physical interpretation behind the result.]
The {\it exact} values of
the numerical coefficients can be  obtained  from simulations
and it changes 14.14 to 11.7 and 200 to 125 in the above relations.

  The result we have obtained for the non linear end corresponds to an assumption called
       {\it stable clustering} \index{stable clustering}. If we ignore the
effect of mergers, then it seems reasonable that virialized systems 
should maintain their densities and sizes in proper coordinates, i.e.
the clustering should be ``stable". This
would require the correlation function to have the form $\bar\xi_{NL}
(a,x)=a^3F(ax)$. [The factor $a^3$ arising from the decrease in
background density]. One can easily show that this will lead to $U \propto
\xi_L^{3/2} $ in the non linear regime. 
Another way deriving this result is to note that if the {\it proper} size of the objects
do not change with time, then $\dot r=0$ which implies, statistically, $v=-\dot a x$
requiring
 $h=1$. Integrating (\ref{xibarint}) with appropriate boundary
condition leads to $U \propto
\xi_L^{3/2} $ .

  In case
mergers of structures are important, one would consider this
assumption to be suspect \cite{tpjpo}. We can,
however, generalize the above 
argument in the following manner: If the virialized systems have
reached  stationarity in the statistical sense, 
then it seems reasonable to assume that
the function $h$
--- which is the ratio between two velocities --- should reach some
constant value. In that case, one can integrate (\ref{xibarint}) and
obtain the result $\bar\xi_{NL}=a^{3h}F(a^hx)$ where $h$ now denotes the asymptotic value. A similar argument
will now show that
\begin{equation}
\bar\xi_{NL}(a,x)\propto [\bar\xi_{L}(a,l)]^{3h/2}\label{qnlscl2}
\end{equation}
in the general case. 
If the linear power spectrum has an index $n$ (with $P_L(k)\propto k^n, \bar \xi_L \propto x^ {-(n+3)}$) then
one would get
\begin{equation}
\bar{\xi}(a,x)\propto a^{(3-\gamma)h}x^{-\gamma};\qquad 
\gamma={3h(n+3)\over 2+h(n+3)}
\end{equation}
Simulations are not accurate enough to fix the value of $h$; in
particular, the asymptotic value of $h$ could depend on $n$
within the accuracy of the simulations.

It is possible to obtain similar relations between $\xi(a, x)$ and
$\xi_L (a, l) $ in two dimensions as well by repeating the above analysis \cite{tpicgc}. 
  In 2-D the scaling
relations turn out to be
\begin{equation}
\bar \xi (a,x)\propto \cases{\bar \xi_L (a,l)&({\rm Linear}) \cr
\bar\xi_L(a,l)^2 &({\rm Quasi-linear})\cr
\bar\xi_L(a,l)^{h} &({Nonlinear}) \cr}
\end{equation}
where $h$ again denotes the asymptotic value. For power law spectrum the nonlinear correction function will
$\bar\xi_{NL} (a, x) = a^{2 - \gamma} x^{-\gamma} $ with $\gamma = 2
(n + 2) / (n + 4)$. 
If we generalize the concept of stable clustering to mean constancy of
$h$ in the nonlinear epoch, then the correlation function will behave
as $\bar\xi_{NL} (a, x) = a^{2h}F(a^hx)$. In this case, if the
spectrum is a power law then the nonlinear and linear indices are
related to 
\begin{equation}
\gamma = {2h (n + 2) \over 2 + h (n + 2)}
\end{equation}
[In general, $\bar\xi \propto \xi_L^D$ in the quasi-linear regime and $\bar\xi \propto \bar\xi_L^{Dh/2}$
in  the nonlinear regime \cite{nktpprd}  where $D$
is the dimension of space]. All the features discussed in the case of 3 dimensions are present
here as well. For example, if the asymptotic value of $h$ scales with
$n$ such that $h (n + 2 )= {\rm constant}$ then the nonlinear index
will be independent of the linear index. Figure \ref{figtwod}   shows the results of numerical simulation in 2D, which suggests that $h= 3/4$ asymptotically \cite{jsbtpsunu}

%%%%%%%%%% Figure %%%%%%%%%%%%%%%%%%%%%%
\begin{figure}[ht]
\begin{center}
\includegraphics[width=.8\textwidth]{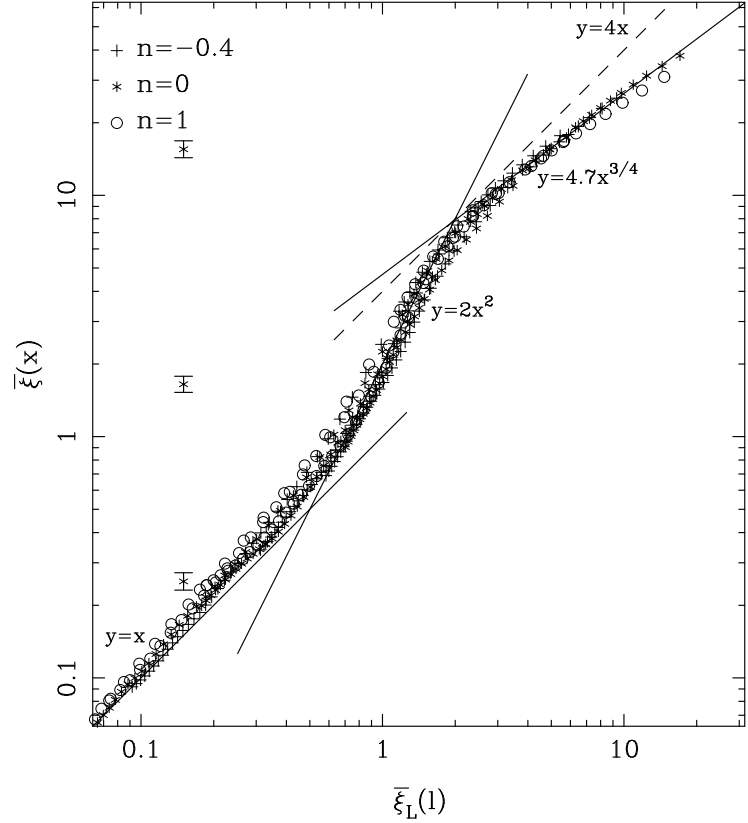}
\end{center}
\caption{
The NSR in 2D giving $\bar\xi(x)$ as a function of the linear mean correlation function
$\bar\xi_L(l)$. The theoretical predictions at the three regimes are shown by 
solid lines of slopes $1, 2, 3/4$. The data points are from
simulations for different power spectra and different epochs. 
The overlapping of data points shows the existence of an NSR and its agreement
with the theoretical prediction lends support to the model discussed in the 
text. The error bars indicate the typical accuracy of the result}
\label{figtwod}
\end{figure}
%%%%%%%%%% End   Figure   %%%%%%%%%%%%%%%%%%%%%%

The ideas presented here can be generalized in two obvious directions \cite{dmtp}: 
(i) By considering peaks of
different heights, drawn from an initial Gaussian random field,
and averaging over the probability distribution one can obtain
a more precise NSR. (ii) By using a generalized ansatz for
higher order correlation functions, one can attempt to compute
different statistical  parameters in the quasi linear and nonlinear regimes. 

  The entire approach side steps three important issues which needs to be
  investigated more thoroughly than has been done in the literature.
\begin{itemize}
 \item 
   There has been no ``first-principle" derivation of the non linear
  scaling relations in spite of several attempts by different groups.
  (The NSR does not depend on $G$, for example!)
  For example, I have not obtained the NSR from the basic
  equation  (\ref{powtransf}) derived earlier in section \ref{gravclnl}  
   because I (or anyone else) do not know how to go about
   doing this.
  \item 
  A strong constraint any correlation function must satisfy is that
   its Fourier transform must be positive definite. The necessary and 
   sufficient condition for a function $f(x)$ to have a positive definite
   Fourier transform is that it must be a convolution, viz., it must be
   expressible as an integral over $y$ of a product $g(x+y) g(y)$
   where $g$ is some other function. In the NSR, one starts with a 
   linear correlation function and maps it by a functional transform
   to obtain a non linear correlation function. There is no guarantee 
   that this functional transform will actually lead to a valid correlation 
   function in the sense that it will have a positive definite Fourier
   transform. In fact, it will not except approximately; it is not clear
   what are the implications of this result. 
   \item
    Both $\xi$ and $\bar\xi$ can be negative for a range of 
   scale. The NSR described above is not applicable to power
   spectra with negative correlation function. It is 
   possible to generalize the formalism to cover this situation
   but such an extension leads to results which are fairly counter intuitive
   \cite{tpnk}.
\end{itemize}

\section{\label{cipt} Critical indices\index{Critical indices} and  power transfer\index{power transfer} in gravitational clustering}

Given a model for the evolution of the power spectra in the quasi linear
and nonlinear regimes, one could  explore whether 
evolution of gravitational clustering
possesses any universal characteristics. 
  The derivation of NSR in the previous section has encoded in it the feature
  that $n=-1$ and $n=-2$ (in the power spectrum with $n$ defined
through the relation $P\propto k^n$) will appear as ``critical indices'' in the quasi-linear
  and non linear regimes respectively, 
  in the sense that $\bar\xi \propto a^2$ for these indices (which matches with 
  the linear evolution, $\xi_L \propto a^2$).

   This suggests even a stronger result:
  any generic initial power spectrum will be driven to a
 particular form of power spectrum in the late stages of the evolution. 
Such a possibility arises because of the following reason: 
We see from equation (\ref{hamilton}) that [in the
quasi linear regime] spectra with $n<-1$ grow faster
than $a^2$ while spectra with $n>-1$ grow slower than $a^2$. This feature
could drive the spectral index to $n=n_c\approx -1$ in the quasi linear
regime irrespective of the initial index. Similarly, the index in
the nonlinear regime could be driven to $n\approx -2$ during the late time evolution. So the spectral indices $-1$ and $-2$ are some kind
of ``fixed points" in the quasi linear and nonlinear regimes.

This effect can be understood better by studying the ``effective" index
for the power spectra at different stages of the evolution    \cite{jsbtp1} .
Let us define a {\it local} 
index for rate of clustering by
\begin{equation}
n_a(a,x)\equiv \part{\ln \xb}{\ln a}
\end{equation}
which measures how fast $\xb$ is growing. When $\xb\ll 1$, then $n_a=2$
irrespective of the spatial variation of $\xb$ and the evolution preserves the shape of $\xb$. However, as clustering develops, the growth rate will
depend on the spatial variation of $\xb$. Defining the effective spatial
slope by
\begin{equation}
-[n_{x}(a,x)+3]\equiv \part{\ln \xb}{\ln x}
\end{equation}
one can rewrite the equation (\ref{qhsi}) as
\begin{equation}
\label{naeqn}
n_a=h(\frac{3}{\xb} -n_{x})
\end{equation}
At any given scale of nonlinearity, decided by $\xb$, there exists a critical
spatial slope $n_c$ such that $n_a>2 $ for $n_{x}<n_c$ [implying rate of growth is faster
than predicted by linear theory] and 
$n_a<2 $ for $n_{x}>n_c$ [with the rate of growth being slower
than predicted by linear theory]. The critical index $n_c$ is fixed by setting $n_a=2$ in  equation (\ref{naeqn}) at any instant.  This requirement is established from the physically motivated desire to have a form of the two point correlation function that remains invariant under time evolution. Since the linear end of the two point correlation function scales as $a^2$, the required invariance of form constrains the index $n_a$ to be $2$ at {\it all} scales . The fact that $n_a>2$ for $n_{x} <n_c$ and $n_a<2$ for $n_{x} >n_c$   will tend  to ``straighten out'' correlation functions  towards the critical slope.
[We are assuming that $\xb$ has a slope that is decreasing with
scale, which is true for any physically interesting case]. From the NSR it is easy to see that in the range $1 {\mbox{\gaprox}} 
\bar\xi {\mbox{\gaprox}} 200$, the critical index is $n_c\approx -1$
and for $200 \gaprox \bar\xi$, the critical index is $n_c\approx -2$. 
This clearly suggests that the local effect of evolution is to
drive the correlation function to have a shape with $(1/x)$ behaviour
at nonlinear regime and $(1/x^2)$ in the intermediate regime. Such a 
correlation function will have $n_a\approx 2$ and hence will grow at
a rate close to $a^2$.  

To go from the scalings in two limits to an actual profile, we can use
some fitting function. 
 One possible interpolation   \cite{tpsunu}  between the two limits is given by:
\begin{equation}
\label{xisolution}
\bar\xi(a,x)=\left(\frac{Ba}{2}\;\left(\sqrt{1+\frac{L}{x}} -1\right)\right)^2
\label{interpol}
\end{equation}
with $L, B$ being constants.   If we evolve this  profile  (with the optimum value of
$B=38.6$)
from $a^2=1$ to $a^2\approx 1000$ using the NSR, and plot 
$[\bar\xi(a,x)/a^2]$ against $x$, then  the curves virtually fall on top of each other within about 10 per cent  \cite{tpsunu}.
 This   shows that the profile does grow 
approximately as 
$a^2$.

These considerations also allow us to predict the nature of power
transfer in gravitational clustering. Suppose that, initially, the
power spectrum was sharply  
peaked at some scale
$k_0=2\pi/L_0$ and has a small width $\Delta k$. When the peak
amplitude of the spectrum is far less than unity, the evolution
will be described by linear theory and there will be no flow
of power to other scales. But once the peak approaches a value
close to unity, power will be generated at other scales due to nonlinear
couplings even though the amplitude of perturbations in
these scales are less than unity.
 For modes with no
initial power [or exponentially small power],  nonlinear
coupling provides the only driving term in (\ref{exev}). These generate power at the scale ${\bf k}$
through mode-coupling, provided power exists at some other scale. 
[We saw this effect in section \ref{gensmall}  for a simple model.]

In ${\bf x}-$space, this arises along the following lines: If the initial
spectrum is sharply peaked at some scale $L_0$, first structures to
form  are voids with a typical diameter
$L_0$. Formation and fragmentation of sheets bounding the voids lead
to generation of power at scales $L<L_0$. First bound structures will then form
at the mass scale corresponding to $L_0$. In such a model, 
$\bar\xi_{\rm{lin}}$ at $L<L_0$  is nearly constant with an effective index of
$n\approx -3$. Assuming we can use equation (\ref{hamilton}) with the
local index in this case, we expect the power to grow very rapidly
as compared to the linear rate of $a^2$. [The rate of growth is $a^6$
for $n= -3$ and $a^4$ for $n=-2.5$.] Different rate of growth for
regions with different local index will lead to steepening of
the power spectrum and an eventual slowing down of the rate of
growth. In this process, which is the dominant one, 
 the power transfer is mostly
from large scales to small scales. [Except for the 
generation of the $k^4$ tail at large scales which we have discussed earlier
in subsection \ref{nltail}.]

%%%%%%%%%% Figure   %%%%%%%%%%%%%%%%%%%%%%
\begin{figure}[ht]
\begin{center}
\includegraphics[width=.8\textwidth]{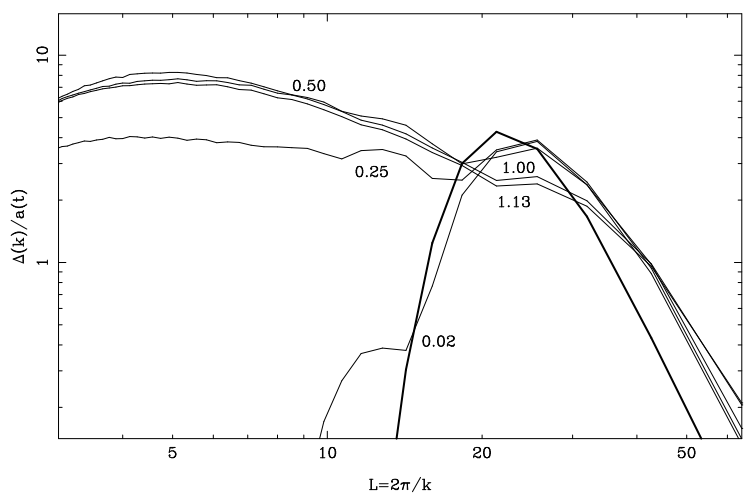}
\end{center}
\caption{The transfer of power in gravitational clustering }
\label{ptgravcl}
\end{figure}
%%%%%%%%%% End  Figure   %%%%%%%%%%%%%%%%%%%%%%

%%%%%%%%%% Figure   %%%%%%%%%%%%%%%%%%%%%%
\begin{figure}[ht]
\begin{center}
\includegraphics[width=.8\textwidth]{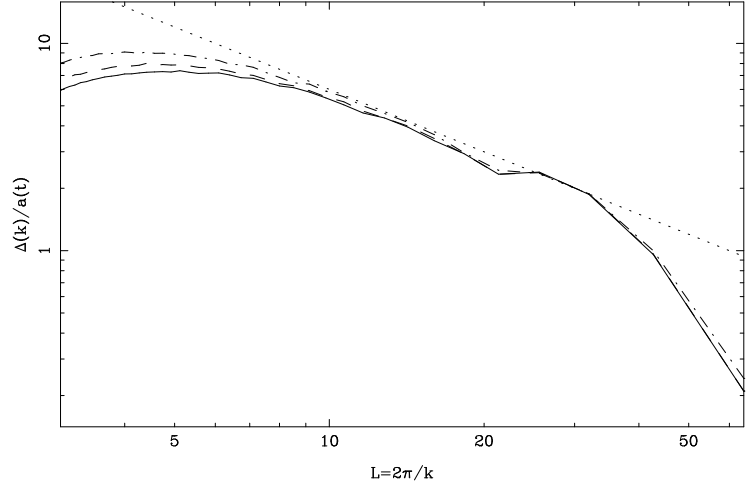}
\end{center}
\caption{The growth of gravitational clustering towards a universal power spectrum $P(k) \propto k^{-1}$}
\label{univpt}
\end{figure}
%%%%%%%%%% End   Figure  %%%%%%%%%%%%%%%%%%%%%%

From our previous discussion, we would have expected such an evolution
to lead to a  ``universal'' 
power spectrum\index{universal
power spectrum} with some critical index $n_c\approx -1$ 
for which the rate of growth is that of linear theory - viz.,
$a^2$. In fact, the same results should  hold even when there exists small
scale power;  numerical simulations  confirm this
prediction and show that - in the quasi linear
regime, with $1<\delta<100$ - power spectrum indeed has a universal slope (see figures \ref{ptgravcl}, 
\ref{univpt}; for more details, see \cite{jsbtp1}).

The initial power spectrum for figure \ref{ptgravcl} was
a Gaussian peaked at the scale $k_0=2\pi/L_0 ; L_0=24$ and having a
spread $\Delta k=2\pi/128$. The amplitude of the peak was chosen so
that $\Delta_{lin} (k_0=2\pi /L_0, a=0.25)=1$, where $\Delta^2(k)=k^3
P(k)/(2\pi^2)$ and $P(k)$ is the power spectrum. 
The y-axis is
$\Delta(k)/a$, the power per logarithmic scale divided by the linear
growth factor. This is plotted as a function of scale $L=2\pi/k$ for
different values of scale factor $a(t)$ and the curves are labeled by the
value of $a$. As we have divided the power spectrum by its linear rate
of growth, the change of shape of the spectrum occurs strictly because
of non-linear mode coupling. It is clear from this figure that power at
small scales grows rapidly and saturates at the  growth rate close to
the linear rate [shown by crowding of curves] at later epochs. The
effective index for the power spectrum approaches $n=-1$ within the
accuracy of the simulations. This figure clearly demonstrates the
general features we expected from our understanding of scaling
relations.

Figure \ref{univpt} compares  power spectra of   three different models at a late epoch. 
Model I was described in the last para;  Model II had initial power concentrated in {\it two} narrow windows in
$k$-space. In addition to power around $L_0=24$ as in model I, we
added power at $k_1=2\pi/L_1 ; L_1=8$ using a Gaussian with same width
as that used in model I. Amplitude at $L_1$ was chosen five times
higher than that at $L_0=24$, thus $\Delta_{lin} (k_1,a=0.05)=1$.
 Model III was similar to model II, with the small scale peak
shifted to $k_1=2\pi/L_1 ; L_1=12$. The amplitude of the small scale
peak was the same as in Model II. At
this epoch $\Delta_{lin}(k_0)=4.5$ and it is clear from this figure
that the power spectra of these models are very similar to one
another.

\section{\label{univgravcl}Universal behaviour of gravitational clustering in the  asymptotic limit\index{asymptotic limit}}

   In the study of isolated gravitating systems, it is easy to arrive at a broad picture
   of the late time structure of the system. We have seen that it will be made of 
   a very compact core and a diffuse halo and ---if  the system is confined by a reflecting
   wall --- then this state could be thought of as being made of two distinct phases 
   as described in section \ref{sec:phases}.   By and large, all memory of 
   initial stage could have been wiped out in the asymptotic steady state.
   
   The situation is much more complicated in the case of gravitational clustering in an
   expanding background. To begin with, defining an asymptotic state in a universe
   which is expanding according to $a(t)\propto t^{2/3}$ itself is problematic.
   Taking a strict $t\to \infty$ limit is meaningless and hence one is interested in 
   time-scales which are finite but much longer than  other time scales in the problem.
   We need to make this notion more precise. Secondly, we saw in section \ref{renormgrav} that
the virialized cores and the diffuse halo are only weakly coupled in the cosmological context. Hence
the memory of initial conditions cannot be wiped out as easily in this case as, for example, in the context of
systems with short range interaction or even in the context of gravitating systems in static backgrounds.
We will now discuss some of these issues.
   
    Let us start with a time $t=t_i$ at which the density fluctuations $\delta_k$
   is much smaller than unity at all scales. (If necessary, we shall assume that there 
   are cut-offs at small and large scales.) The evolution in the initial stages can then be 
   understood by linear theory developed in section \ref{evollarge} ; the power at all scales
   grow as $a^2$ and the power spectrum maintains its shape. As time goes on, the density
   contrast at some scale will hit unity and structures corresponding to that characteristic scale
   will begin to get formed. If the power spectrum decreases with $k^{-1}$ (so that there is more
   power at small spatial scales), then small scales will go non linear first. 
   
   Once the non linear
   structure has formed, we need to study the evolution at different wave numbers differently.
   If $k_{\rm nl}$ is the scale at which $\delta \approx 1$, then our analysis in section \ref{renormgrav} 
   shows that the evolution is still linear at $k\ll k_{\rm nl}$ provided the initial power was 
   more than $k^4$; if not, a $k^4 $ tail develops rapidly at these large spatial scales. At
   $k \gg k_{\rm nl}$, virialized structures would have formed, which are fairly immune to
   overall expansion of the universe. The discussion in section \ref{renormgrav} shows   that 
   these scales also do not affect the evolution at smaller $k$. In this sense, the universe
   decouples nicely into a clustered component and an unclustered one which do not affect
   each other to the lowest order. The situation is most complex for $k\approx k_{\rm nl}$ since
   neither of the approximations are possible. 
   
   The critical scale $k_{\rm nl}$ itself is a function of time and, in the popular cosmological
   models, keeps decreasing with time; that is, larger and larger spatial scales go non linear
   as time goes on. In such an intrinsically time dependent situation, one could ask for
   different kinds of universal behaviour and it is important to distinguish between them.  
   
   To begin with, one could examine whether the power spectrum (or the correlation function) 
   has a universal shape and evolution at late times, independent of initial power spectrum.
   We have seen in section \ref{cipt}  that the late time power spectrum {\it does}
   have a universal behaviour if the initial spectrum was sharply peaked. In this case,
   at length scales smaller than the initial scale at which the power is injected,
   one obtains two critical indices $n=-1$ and $-2$ and the two point correlation 
   function has an approximate shape of (\ref{interpol}) or even simpler, 
    $\bar \xi (a,x) \propto a^2x^{-1}(l+x)^{-1}$
   where $l$ is the length scale at which $\bar\xi \approx 200$.
   At scales bigger than the scale at which power was originally injected,
   the spectrum develops a $k^4$ tail.   
   
   But if the initial spectrum is {\it not} sharply peaked, each band of power evolves by this
   rule and the final result is a lot messier.  
   The NSR developed in section \ref{nlscales}  allows one to tackle this 
   situation and ask how the non linear scales will behave. Here 
   one of the most popular assumptions used in the literature is that of stable
   clustering which requires 
         $v=-\dot a x$ (or  $h=1$) for 
sufficiently large $\bar\xi(a,x)$.
Integrating equation (\ref{xibarint}) with $h=1$, we get $\bar\xi(a,x)=\bar\xi_L^{3/2}(a,x)$. 
If $P_{in}(k)\propto k^n$ so that $\bar\xi_L(a,x)\propto a^2 
x^{-(n+3)}$, then $\xb$ at nonlinear  scales will vary as
\begin{equation}
\bar\xi(a,x) \propto a^{\frac{6}{n+5}} x^{-\frac{3(n+3)}{n+5}};\qquad (\bar\xi 
\gg 200)
\end{equation}
if stable clustering is true. Clearly, the power law index in the nonlinear 
regime ``remembers''
the initial index. The same result holds for more general initial conditions.
{\it If} stable clustering {\it is} valid, then the late time  behaviour of $\xb$ is
strongly dependent on the  initial conditions. The two (apparently reasonable) requirements:
(i) validity of stable clustering at highly nonlinear scales and
(ii) the independence of late time behaviour from initial conditions, 
{\it are mutually
exclusive}. 

This is yet another peculiarity of gravity in the context of expanding background. The initial
conditions are {\it not} forgotten --- unlike in systems with short range interactions or even
in the context of gravitating systems in static background. The physical reason for this 
feature is the weak coupling between the virialized ``cores'' and diffuse ``halos'', described in 
section \ref{renormgrav}. This weakness of the coupling keeps the memory of the initial
conditions alive as the system evolves.

  While this conclusion is most probably correct, the assumption of stable clustering
  may not be true. In fact, there is some evidence from simulations that this assumption is not true \cite{tpjpo}. In that context, another
  natural assumption which could replace stable clustering, will be the following:
{\it We assume that $h$ reaches a 
constant
value asymptotically which is not necessarily unity}. Then we get   $\xb=a^{3h}F[a^h x]$ where $h$ now
denotes the constant asymptotic value of of the function. For an initial
spectrum which is scale-free power law with index $n$, this result translates
to 
\begin{equation}
\bar\xi(a,x)\propto a^{\frac{2 \gamma}{n+3}} x^{-\gamma}
\end{equation} 
where $\gamma$ is given by 
\begin{equation}
\gamma=\frac{3 h (n+3)}{2+h(n+3)}
\label{withh}
\end{equation}
One can obtain
a $\gamma$  which is independent of initial power law index provided
$h$ satisfies the condition $h(n+3)=c$, a constant.  Unfortunately, simulations are not good enough yet
to test this conclusion in the really asymptotic domain.

  The second question one could ask, concerns the density profiles of individual
  virialized halos which, of course, is related to the behaviour of the correlation 
  function \cite{tpsunu}, \cite{kandu}, \cite{NFW}.  To focus on this relation, 
   let us start with the simple assumption that the density field $\rho(a,{\bf x})$ at late stages  can 
be expressed as a superposition
of several halos, each with some density profile; that is, we take
\begin{equation}
\label{haloes}
\rho(a,{\bf x})=\sum_{i} f({\bf x}-{\bf  x}_i,a)
\end{equation}
where the $i$-th halo is centered at ${\bf x}_i$ and contributes
an amount $f({\bf x}-{\bf  x}_i,a)$  at the location ${\bf x}_i$  [We can easily generalize this equation to the situation in which there are halos with
different properties, like core radius, mass etc by summing over the number
density of objects with particular properties; we shall not bother to
do this. At the other extreme, the exact description merely corresponds to taking
the $f$'s to be Dirac delta functions. Hence there is no loss of generality in (\ref{haloes})]. The power spectrum for the 
density contrast, $\delta(a,{\bf x})=(\rho/\rho_b-1)$, corresponding to the 
$\rho(a,{\bf x})$ in (\ref{haloes})  can be expressed as
\begin{equation}
P({\bf k},a) = \left( a^3 \left| f({\bf k},a)\right| \right)^2 \left| 
\sum_i \exp -i {\bf k}\cdot{\bf x}_i(a) \right|^2   
= \left( a^3 \left| f({\bf k},a)\right| \right)^2 P_{\rm cent}({\bf 
k},a)
\label{powcen1}
\end{equation}
 where $P_{\rm cent}({\bf k},a)$
denotes the power spectrum of the distribution of centers of the halos.
 This equation (\ref{powcen1}) can be used  directly 
     to probe the nature of  halo profiles\index{halo profiles} along the following lines. If the correlation function varies as
     $\bar\xi \propto x^{-\gamma}$, the correlation function of the centers vary as 
     $\bar\xi \propto x^{-\gamma_C}$ and the individual profiles are of the form 
     $f(x) \propto x^{-\epsilon}$, then the relation 
     $P(k) =|f(k)|^2P_{\rm cent}(k)$ translates to 
     \begin{equation}
     \epsilon = 3+{1\over 2}(\gamma - \gamma_C)
     \label{ecnfw}
     \end{equation}
     At very non linear scales, the centers of the virialized clusters will coincide with 
     the deep minima of the gravitational potential. Hence the power spectrum of the 
     centers will be proportional to the power spectrum of the gravitational potential 
     $P_\phi(k) \propto k^{n-4}$ if $P(k) \propto k^n$. Since the correlation functions
     vary as $x^{-(\alpha + 3)}$ when the power spectra vary as $k^\alpha$, it 
     follows that $\gamma = \gamma_C -4$. Substituting into (\ref{ecnfw}) we find
     that $\epsilon =1$ at the extreme non linear scales. On the other hand, in the 
     quasi-linear regime, reasonably large density regions will act as cluster
     centers and hence one would expect $P_{\rm cent}(k)$ and $P(k)$ to scale 
     in a similar fashion. This leads to $\gamma \approx \gamma_C$, giving
     $\epsilon \approx 3$. So we would expect the halo profile to vary as $x^{-1}$ 
     at small scales steepening to $x^{-3}$ at large scales. A simple interpolation
     for such a density profile will  be 
     \begin{equation}
     f(x) \propto {1\over x(x+l)^2}
     \end{equation}
     Such a profile, usually called NFW profile \cite{NFW}, is often used in cosmology
     though I have not come across the simple theoretical argument given
     above in the literature.
Unfortunately, it is not possible to get this result from more detailed and transparent arguments.

  In general, at very nonlinear scales, the correlation function probes the profile
  of individual halos and $P_{cent}\approx$ constant implying $\gamma_C=3$ in
equation (\ref{ecnfw}). At these scales, we get
\begin{equation}
\label{gammep}
\gamma=2\epsilon-3
\end{equation}
For the situation considered in equation (\ref{withh}), with $h(n+3)=c$,
the
halo profile will have the index
\begin{equation}
\epsilon=3\left( \frac{c+1}{c+2} \right) 
\end{equation}
which is independent of initial conditions.
Note that we are now demanding the asymptotic value of $h$ to {\it explicitly 
depend} on the initial conditions though the {\it spatial} dependence of $\xb$ 
does not.
In other words, the velocity distribution --- which is related to $h$ --- still 
``remembers'' the initial
conditions. This is indirectly reflected in the fact that the growth
of $\xb$ --- represented by $a^{6c/((2+c)(n+3))}$ --- does depend on the
index $n$.

As an example of the power of such a --- seemingly simple --- analysis note the 
following: Since $c \geq 0 $, it follows that $\epsilon > (3/2)$; invariant 
profiles
with shallower indices (for e.g with $\epsilon=1$) discussed above are not consistent 
with the evolution described above.

While the above arguments are suggestive, they are far from conclusive. It
is, however, clear from the above analysis and it is not easy to provide
{\it unique} theoretical reasoning regarding the shapes of the halos. 
The situation gets more complicated if we include the fact that all halos
will not all have the same mass, core radius etc and we have to modify our
equations by integrating over the abundance of halos with a given value of
mass, core radius etc. This brings in more ambiguities and depending on
the assumptions we make for each of these components
and the final results have no real significance. The issue is theoretically wide open.

  \medskip
       
     \section*{Acknowledgement}
     
      I thank D. Lynden-Bell, R. Nityananda, J. P. Ostriker and K. Subramaniam for several
     discussions over the years. I thank organisers of the LesHouches School, especially Thierry Dauxois,
 for inviting me to the Les Houches
     school and for making this contribution. 
     
     \section*{Appendix A: Basic Cosmology}
     
     I summarize in this appendix some of the key equations in cosmology which
     are used in the review. More detailed discussions can be found in standard
     text books  \cite{cosmotext}. 
     
     The cosmological model 
     which we use in our discussion is described by a spacetime interval of the 
     form 
     \begin{equation}
     ds^2 = dt^2 - a^2(t) d{\bf x}^2
     \label{cosone}
    \end{equation}
    where $a(t)$ is called the  {\it expansion factor}\index{expansion factor} which describes the rate of expansion of the universe
     and the units are chosen so that $c=1$. The form of $a(t)$ is determined by the  energy density
      present in the universe and is determined through the equations
     \begin{equation}
     {1\over 2} \dot a^2 - {G\over a} \left( {4\pi \over 3} \rho_b a^3\right) =0; \quad d(\rho_b a^3) = - p da^3
     \label{costwo}
     \end{equation}
     where $\rho_b $ is the energy density of matter and $p$ is the pressure  related to $\rho_b $  by
     an equation of state of the form $p = p(\rho_b)$. This equation of state, along with the second
     equation in (\ref{costwo}) determines $\rho_b$ as a function of $a$. Substituting in the 
     first equation in (\ref{costwo}), one can determine $a(t)$. As a mnemonic (and only as
     a mnemonic), one can think of 
     the first equation as giving the sum of kinetic energy and potential energy of the universe to  be 
     zero for the universe and the second equation as an equivalent of $dE = - pdV$.
     There are also equivalent to the relation $(\ddot a/a) = -(4\pi G/3)\rho_b(t)$
     
     Non relativistic matter moving  at speeds far less than speed of light ($v\ll c$) will have
     the pressure $p\approx \rho_b v^2$ negligibly small compared to the energy density 
     $\rho_b c^2$. We can then set $p\approx 0$ obtaining $\rho_b \propto a^{-3}$ and
     $a\propto t^{2/3}$. Since an overall multiplication constant in $a$ can be absorbed
     by rescaling the lengths, the normalization of $a(t)$ is arbitrary. It is conventional
     to take $a=(t/t_0)^{2/3}$ with $t_0$ denoting the current age of the universe.
     The rate of expansion at present is called the Hubble constant and is defined by
     \begin{equation}
     H_0 = \left( {\dot a\over a}\right)_0 = {2\over 3t_0}
     \end{equation}
     Observationally, $H_0 \approx 75$ km s$^{-1}$ Mpc$^{-1}$.
     This defines a natural time scale $t_H= H_0^{-1} \approx 10^{10}$ years and a length scale
     $d_H =c H_0^{-1} \approx 4000$ Mpc. 
     Note that $H_0$ and the energy density are related by $\rho_0 = (3 H_0^2/8\pi G)$.
     
     The general relativistic effects of gravity are felt
     over length and time scales comparable to $t_H, d_H$. Since the typical separation between
     galaxies is about 1 Mpc and the size of large superclusters in the universe is about
     100 Mpc, the general relativistic effects are not important at the present epoch for 
     most of the issues in structure formation.           
     Equation (\ref{cosone})  then suggests that if we use the coordinate ${\bf r} \equiv  a(t) {\bf x}$,
     the physical laws in an expanding universe will take more familiar form at small
     scales. (The ${\bf r}$ is called the {\it proper coordinate}, while ${\bf x}$ is called the
     {\it comoving coordinate}.)
     For example, gravity can be described by Newtonian theory in proper coordinates
     with the cosmic background providing an extra   gravitational potential 
     $\Phi_{\rm FRW} = - (2\pi G/3) \rho_b(t) r^2$
     corresponding to a uniform distribution of matter. 
     
     When matter expands with the universe homogeneously,
       the proper
     coordinate separation between any two points vary as $\dot {\bf r} = \dot a {\bf x} = (\dot a/a){\bf r}$.
     (This is called Hubble expansion with ${\bf v}=H{\bf r}$.)
     In this case, 
     \begin{equation}
     \ddot{\bf r} = - {\ddot a\over a} {\bf r} = {4\pi G\over 3} \rho_b(t) {\bf r} = - \nabla_{\bf r} \Phi_{\rm FRW}
     \label{cosacc}
     \end{equation}
     Gravitational clustering and growth of inhomogeneities require particles to move relative to the 
     cosmic expansion with $\dot {\bf x} \ne 0$. 
     
     In the study of structure formation, the central quantity one uses is the density contrast
     defined as $\delta (t, {\bf x}) = [\rho(t,{\bf x}) - \rho_b(t)]/\rho_b(t)$ which characterizes 
     the fractional change in the energy density compared to the background. 
     (Here $\rho_b(t) = <\rho(t,{\bf x})>$ is the mean background density.)
     Since one is 
     often interested in the statistical behaviour of structures in the universe, it is conventional
     to assume that $\delta$ and other related quantities are  elements of an ensemble.
     Many popular models of structure formation suggest that the initial density perturbations
     in the early universe can be represented as a Gaussian random variable with zero mean
     (that is, $<\delta> =0$) and a given initial power spectrum. The latter quantity is defined through
     the relation $P(t, k) = <|\delta_k(t)|^2>$ where $\delta_{\bf k}$ is the Fourier transform
     of $\delta(t,{\bf x})$. It is also conventional to define the two-point correlation function
     $\xi(t,x)$   
     and the mean correlation function $\bar\xi(t,x)$ via the equations
      \begin{equation}
      \xi(t,x) = \int {d^3 k\over (2\pi)} \, P(t,k) \, e^{i{\bf k\cdot x}}; \quad \bar \xi(t,x) = {3\over x^2} 
      \int_0^x \xi(t,y)\, y^2 dy
      \end{equation}     
     Though gravitational clustering will make the density contrast non Gaussian at late times,
     the power spectrum and the correlation functions continue to be of primary importance
     in the study of structure formation.

 \end{document}